\documentclass[twocolumn]{aastex631}
\pdfoutput=1
\usepackage{graphicx}
\usepackage{bm}
\usepackage{amssymb}
\usepackage{amsmath}
\usepackage{units}
\usepackage{array}
\usepackage{hyperref}
\usepackage{soul}
\usepackage{placeins}
\usepackage{float}
\usepackage{multirow}
\usepackage{makecell}

\usepackage[usestackEOL]{stackengine}
\edef\tmp{\the\baselineskip}
\setstackgap{L}{\tmp}

\def\fref{Fig.~\ref}
\def\sref{\S \ref}

\begin{document}

\title{Eccentric Minidisks in Accreting Binaries}
\shorttitle{Eccentric Minidisks in Accreting Binaries}

\correspondingauthor{John Ryan Westernacher-Schneider}
\email{john.westernacher.schneider@gmail.com}

\author[0000-0002-3047-7200]{John Ryan Westernacher-Schneider}
\affiliation{Leiden Observatory, Leiden University, P.O. Box 9513, 2300 RA Leiden, The Netherlands}
\affiliation{Department of Physics and Astronomy, Clemson University, Clemson, SC 29634, USA}

\author[0000-0002-1895-6516]{Jonathan Zrake}
\affiliation{Department of Physics and Astronomy, Clemson University, Clemson, SC 29634, USA}

\author[0000-0002-0106-9013]{Andrew MacFadyen}
\affiliation{Center for Cosmology and Particle Physics, Physics Department, New York University, New York, NY 10003, USA}

\author[0000-0003-3633-5403]{Zolt\'an Haiman}
\affiliation{Department of Astronomy, Columbia University, New York, NY 10027, USA}


\begin{abstract}
We show that gas disks around the components of an orbiting binary system (so-called minidisks) may be susceptible to a resonant instability which causes the minidisks to become significantly eccentric. Eccentricity is injected by, and also induces, regular impacts between the minidisks at roughly the orbital period of the binary.
Such eccentric minidisks are seen in vertically integrated, two-dimensional simulations of a circular, equal-mass binary accreting from a circumbinary gas disk with a $\Gamma$-law equation of state. Minidisk eccentricity is suppressed by the use of an isothermal equation of state.
However, the instability still operates, and can be revealed in a minimal disk-binary simulation by removing the circumbinary disk, and feeding the minidisks from the component positions. Minidisk eccentricity is also suppressed when the gravitational softening length is large ($\gtrsim 4\%$ of the binary semi-major axis), suggesting that its absence could be an artifact of widely adopted numerical approximations; a follow-up study in three dimensions with well-resolved, geometrically thin minidisks (aspect ratios $\lesssim 0.02$) may be needed to assess whether eccentric minidisks can occur in real astrophysical environments. If they can, the electromagnetic signature may be important for discriminating between binary and single black hole scenarios for quasi-periodic oscillations in active galactic nuclei; in turn, this might aid in targeted searches with pulsar timing arrays for individual supermassive black hole binary sources of low-frequency gravitational waves.
\end{abstract}


\section{Introduction}

\begin{figure*}[ht]
\centering
\includegraphics[width=1\textwidth]{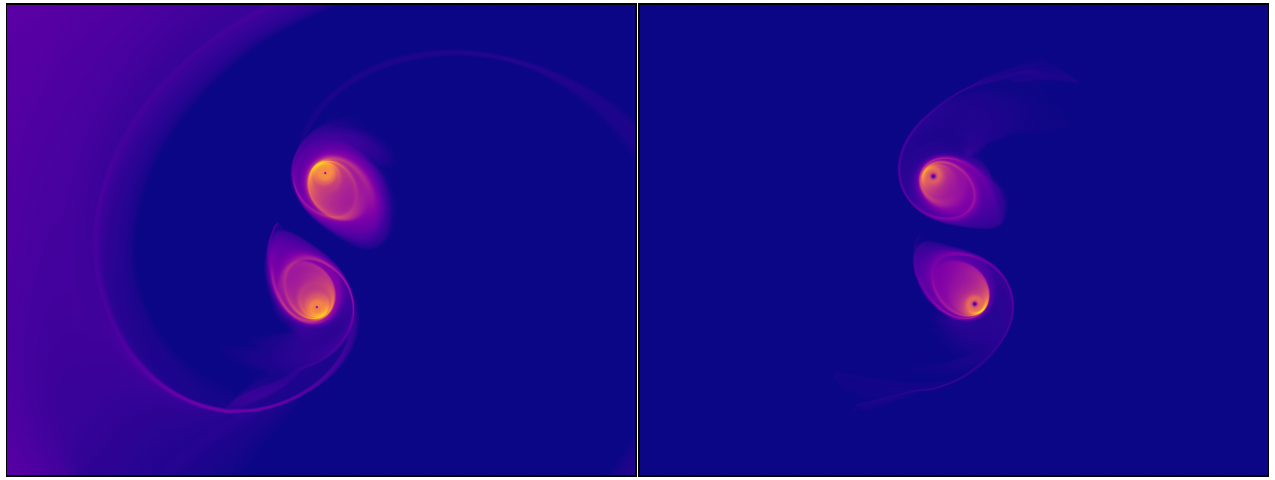}
\caption{Snapshots of $\sqrt{\Sigma}$ from two of our high- and very high-resolution runs, on independent color scales. Left: run labeled VH3 with resolution $\Delta x=0.00125\,a$ (see Table~\ref{tab:simsuite}). Right: the corresponding depletion experiment with resolution $\Delta x=0.0025\,a$, where the circumbinary disk has been removed (run labeled H2).} \label{fig:fig1}
\end{figure*}

There is strong evidence that some galaxies host supermassive black hole binaries at their centers \cite[SMBHBs;][]{begelman+1980}.
These objects are powerful sources of low-frequency gravitational wave (GW) radiation, and their population has long been theorized to generate a stochastic GW background \citep{Carr1980, Rajagopal1995}. The recent detection of such a background with pulsar timing arrays \citep{Agazie2023, antoniadis+2023a, xu+2023, reardon+2023a}, and indications that the SMBHB population is indeed its likely source \citep{Afzal2023, antoniadis+2023b, reardon+2023b}, creates a new imperative to identify individual SMBHB systems.

Galaxies that host SMBHBs are likely to be active, due to the presence of copious circumnuclear gas associated with the galaxy merger that gave rise to the black hole pair \citep{Barnes1992}. Such binary AGNs may exhibit quasi-periodic oscillations (QPOs) connected in some way to the binary's orbital motion, and more than 200 binary AGN candidates have been proposed based on QPO detections in electromagnetic (EM) surveys \citep[e.g.][]{graham+2015, charisi+2016, liu+2019, liu+2020, chen+2020, penil+2022, chen+2022, Li2023}. Those SMBHB candidates, and others yet to be discovered, could become joint GW-EM sources as the pulsar timing arrays continue to accumulate sensitivity.

The identification of an individual low-frequency GW source with a particular time-varying AGN will be challenging, firstly, because of theoretical uncertainties in the relationship between the binary's GW and EM temporal signatures, and secondly, because
AGN periodicity can also signify processes that have nothing to do with a binary, such as jet precession or accretion disk instabilities \cite[see e.g.][for a review]{Hawkins2002}. It means that many of the binary AGN candidates so far identified could be single accreting SMBHBs, exhibiting real periodicities. To separate the single black hole AGNs from the binaries will thus require a theoretical understanding of the unique temporal characteristics of the binary accretion process.

In a black hole binary accretion system, gas comes to the binary through a circumbinary disk (CBD), which feeds gas into minidisks around each of the black holes. The electromagnetic emission from the minidisks and CBD could be significantly affected by their dynamical interaction with the binary, leading to an imprint on many bands across the electromagnetic spectrum \citep[e.g.][]{sesana+2012, farris+2015b, Panessa2019}.

Variability in $\gamma$-rays is also expected in binary blazars, due to modulation of the rates of mass delivery to the black holes \citep{Sobacchi2017, penil+2020}. Computational studies have revealed a variety of mechanisms that could lead to detectable, periodically varying EM output, associated specifically with the dynamics of binary accretion \citep[e.g.][]{Komossa2006, macfadyen+2008, noble+2012, Roedig+2012, shi+2015, Dorazio+2016, Munoz2019, duffel:2020:massratio, Zrake:2021:eccentric, franchini+2023}. 
These range from periodic variations at or near the orbital period, to those over tens of binary orbits, and they originate from distinct spatial regions of the accretion system.

In a previous study \citep[][henceforth W22]{WS2022}, we reported multi-wavelength light curves of thermal emission from accreting black hole binaries, computed from vertically integrated 2-dimensional viscous hydrodynamics simulations with a detailed treatment of the radiative cooling. In that paper, we identified a strong QPO in the disk thermal luminosity at roughly the binary orbital period. However, we did not give a detailed discussion of the physical origin of the QPO, nor whether the same one had been identified previously by other authors. In this paper we report on the physical mechanism of the QPO we found in W22, and present very high resolution simulations revealing that it arises from an instability in which the minidisks become eccentric and exchange mass at a regular interval -- see \fref{fig:fig1} for sample snapshots. The instability is sensitive to the thermodynamic treatment, being generally suppressed when locally isothermal or target-temperature ``$\beta$-cooling'' approximations are used.

Aside from the need to understand the temporal structure of EM output from accreting binaries to identify SMBHBs among candidates, there is also a new imperative to understand CBD morphology as radio imaging (at especially high resolution by ALMA) has revealed dramatic examples of well-resolved substructures in proto-planetary disks including those around young stellar binaries \citep[e.g.][]{andrews+2018, huang+2020, sierra+2021}.

Our paper is organized as follows. In \sref{sec:background} we review theoretical results on periodic light curves from accreting binaries. In \sref{sec:models} we describe the simulation setup and diagnostics used to quantify the minidisk morphology. We describe the numerical approach in \sref{sec:numerics}. Our simulation results are reported in \sref{sec:results}, including subsections on the irrelevance of infall from the CBD for driving minidisk eccentricity (\sref{sec:CBDinfall}), the central role of the minidisk-minidisk interaction (\sref{sec:interaction}), the periodic mass trade between minidisks and precession effects (\sref{sec:masstrade}), dependence on various parameters and prescriptions (\sref{sec:depsoft}-\sref{sec:dephole}), numerical convergence of the minidisk eccentricity growth rate (\sref{sec:convergence}), and a summary of our key numerical findings (\sref{sec:results-summary}). We provide discussions in \sref{sec:discuss} on the mechanism of the eccentric instability (\sref{sec:mechanism}), a comparison to other eccentricity-driving mechanisms (\sref{sec:other}), the role of gravitational softening (\sref{sec:softening} \& \sref{sec:precess}), guidance for future 3-dimensional studies (\sref{sec:pars}), and the observable consequences of the minidisk mass trade (\sref{sec:obs}). We conclude in \sref{sec:conclude}, and our suite of simulations is tabulated in the appendix.

\section{Background} \label{sec:background}
We briefly review some of the mechanisms known to cause periodic oscillations in the light curves of accreting binary systems. Several of these are related to the formation of an $m=1$ over-density in the CBD, referred to here as the ``lump'' \citep[e.g.][]{macfadyen+2008}.\footnote{Note that, although \cite{macfadyen+2008} observed the dominant $m=1$ over-density in the viscously relaxed CBD state, and its periodic imprint on the mass accretion rate into the cavity, the terminology of ``lump'' was coined later.} The lump forms near the inner edge of the CBD at $r \gtrsim 3a$, and orbits the binary at roughly the Kepler frequency (for reference that is $\sim 5$ binary orbits if the lump orbits at a radius of $3a$; in general we denote the lump orbital frequency as $f_{\rm lump}$). Its presence leads to a modulation in the rate of mass delivered from the CBD to the minidisks on the time scale $1/f_{\rm lump}$, because gas orbits in the CBD are generally eccentric, and feeding to the binary is enhanced when the lump orbits through its closest approach to the binary.

Variations in the rate of feeding to the minidisks are not necessarily transmitted to the black holes. Indeed, the time scale for mass to accrete through the minidisks is typically in the range of 10's of binary orbits. However, simulations by a number of authors \cite[e.g.][]{Farris+2014, Zrake:2021:eccentric} indicate that in spite of possible buffering by the minidisks, the $5-10$ orbit lump period can still be detectable at the $\sim 10\%$ level in the time series of the accretion power. Also, gas injection to a minidisk involves the impacts of gas streams from the CBD, and the resulting disturbances may propagate to the black hole in a fraction of an orbit, much faster than the viscous rate. Furthermore, in radiatively efficient environments, the EM light curves could reflect the stream-minidisk impacts even if the black holes themselves accrete steadily. QPOs associated with lump-induced variation of gas delivery from the CBD may thus be a detectable feature of binary AGN light curves.

For a circular, equal-mass binary, the presence of the lump can lead to a second kind of periodic oscillation, at the frequency $2(f_{\rm bin} - f_{\rm lump})$ \citep[e.g.][]{noble+2012}. This ``binary-lump beat frequency'' is the frequency at which one black hole or the other overtakes the orbiting lump. Simulations indicate that mass delivery to the binary is modulated at this frequency when the CBD extends inwards far enough for tidal stripping of the CBD to operate at any orbital phase, and thus be enhanced any time a binary component passes the lump. When the low-density cavity around the binary is very large, gas is only tidally stripped from the near side of the eccentric CBD, and the enhanced feeding rate at the frequency $2(f_{\rm bin} - f_{\rm lump})$ is suppressed \citep[see examples in][and W22]{farris+2015b} (one could say the duty cycle of this mode is reduced to a fraction of the lump orbit around its closest approach to the binary). For reference, when the lump period is 5 binary orbits, the binary-lump beating operates at a frequency of 1.6 times the binary orbital frequency.

In W22, we computed light-curves of thermal disk emission from an equal-mass black hole binary, and reported a QPO operating at between 1 and 2 times the binary orbital frequency. The similarity in frequency made it easy to confuse this feature with the binary-lump beating effect, however there was an important difference to suggest it had a distinct physical origin: the QPO from W22 showed sensitivity to the length scale parameter $r_{\rm soft}$ used in the code to soften the gravitational potential. In particular, the W22 frequency seemed to approach the orbital frequency as $r_{\rm soft}$ was decreased. The binary-lump beating involves a coupling between the outer edges of the minidisks and the inner edge of the CBD, so it should not be sensitive to how gravity is numerically modeled very near the black holes.

We have carried out a detailed analysis since the publication of W22, and confirmed that indeed the QPO we saw there had nothing to do with the lump, nor the CBD in any direct way. Instead, we found the effect arises due to the minidisks developing a significant eccentricity, and experiencing regular collisions with one another as a result. The minidisks have opposing eccentricity vectors, and the disks collide to produce an EM flare when the long ends of the disks strike one another. The minidisk eccentricity vectors undergo retrograde precession, and the collisions occur at the beat frequency between the minidisk precession and the binary orbit. As we show below, the rate of the minidisk precession increases with $r_{\rm soft}$, and this accounts for the observation from W22 that the eccentric minidisk beat frequency approaches the orbital frequency as softening is decreased. In the subsequent sections we present a detailed characterization of the eccentric minidisk beating effect, and an investigation of the conditions that lead to the growth of minidisk eccentricity.

%
%
\section{Numerical setup} \label{sec:models}
Following W22, we study a binary with total mass $M=M_1 + M_2 = 8\times10^6 M_\odot$ and semi-major axis $a\simeq 10^{-3}\,$pc, yielding an orbital period $T_{\rm bin} \simeq 1\,$yr, but in this paper we consider only an equal mass binary (mass ratio $q=1$) on a circular orbit ($e=0$). The vertically integrated pressure and density are $\mathcal{P}$ and $\Sigma$, respectively, and $\epsilon$ is the specific internal energy. We use an adiabatic equation of state $\mathcal{P}=\Sigma \epsilon (\Gamma -1)$ with $\Gamma=5/3$, neglecting radiation pressure.
In order to obtain numerically tractable Mach numbers $\simeq \mathcal{O}(10)$ while neglecting radiation pressure, the accretion rate must often be extremely super-Eddington \citep[see e.g.][]{dorazio+2013, WS2022}. In this work, gas Mach numbers are chosen in the initial conditions, and are subsequently determined self-consistently, and are typically in the range $\sim 7 - 25$.

We compare constant-$\alpha$ and constant-$\nu$ viscosity models, where $\nu = \alpha c_s h$ is the kinematic viscosity, $c_s = \sqrt{\Gamma\mathcal{P}/\Sigma}$ is the sound speed, $h=\sqrt{\mathcal{P}/\Sigma} / \tilde{\Omega}$ is the disk scale height, and $\tilde{\Omega} \equiv \sqrt{GM_1/r_1^3 + GM_2/r_2^3}$ is a frequency scale accounting for both masses $M_1$, $M_2$ and distances to them $r_1$, $r_2$. We also compare two cooling models, a physical optically thick radiative cooling $\dot{Q} = - (8/3)\sigma T^4 / (\kappa \Sigma)$ where $T$ is the midplane temperature and $\kappa = 0.4\,$cm$^2$/g is the electron scattering opacity, as well as a popular phenomenological target-temperature cooling \citep[e.g.][]{rice+2011} \citep[also known as ``$\beta$-cooling,'' e.g.][]{lin+2016, lyra+2016, boss2017, tartenas+2020, muley+2021, wang+2023, cimerman+2023} prescription $\dot{Q} = - \tilde{\Omega} \Sigma (\epsilon - \epsilon_0) / \beta$ where $\beta$ is a dimensionless parameter, $\epsilon$ is the specific internal energy, and $\epsilon_0=-\Phi / \left[\mathcal{M}^2 \Gamma (\Gamma -1)\right]$ is a target specific internal energy profile, where $\mathcal{M}=10$ is the target orbital Mach number and $\Phi$ is the gravitational potential. $\Omega_K \equiv \sqrt{GM/r_s^3}$ is a softened Keplerian frequency ($r_s \equiv \sqrt{r^2+r_{\rm soft}^2}$ is the softened radial coordinate with $r_{\rm soft}$ the softening length scale). Although our target-temperature cooling models both heat and cool (i.e.~$\dot{Q}=0 \iff \epsilon = \epsilon_0$), we explored a variant which only cools ($\dot{Q}=0$ when $\epsilon<\epsilon_0$) and our conclusions were unaffected.

In addition to cooling models, we also compare our results with locally isothermal runs, where the prescribed sound speed profile corresponds to a uniform orbital Mach number of $10$. This is consistent with the target $\epsilon_0$ we use in our target-temperature cooling runs.

Disk initial conditions correspond to near-equilibrium configurations about a single gravitating object. The gas configuration is allowed to settle around the orbiting binary over several viscous times before analysis begins. This corresponds to $\sim1000-3000$ binary orbits, depending on the model -- see Table~\ref{tab:simsuite}.
The constant-$\alpha$ models exclusively use self-consistent radiative cooling, and their initial conditions are $\Sigma \propto r^{-3/5}$ and $\mathcal{P}\propto r^{-3/2}$.
The constant-$\nu$ models initially have $\Sigma =\,$constant, corresponding to a spatially uniform accretion rate. The subset of constant-$\nu$ models with radiative cooling initially have $\mathcal{P}\propto r^{-3/4}$, corresponding to local balance of viscous heating and radiative cooling, and yielding a Mach number profile $\mathcal{M}\propto r^{-1/8}$. The subset of constant-$\nu$ models with target-temperature cooling instead initially have uniform $\mathcal{M}=10$, which is a popular Mach profile used in both isothermal and target-temperature cooled models, and the target-temperature cooling term drives towards this initial Mach number.

As in W22, black holes are represented by torque-free sinks \citep[see][]{dempsey+2020, Dittmann+2021} with radius $r_{\rm sink}$ and sink rate $s$. Our gravity model derives from a Plummer potential $\Phi_{\rm P} \propto (r^2 + r_{\rm soft}^2)^{-1/2}$. As recognized in the literature \citep[e.g.][]{hure+2009, Mueller+2012}, although some type of softening is numerically necessary to regulate the divergence at a Newtonian point mass, in two-dimensional calculations it physically represents the vertical integration of the force of gravity when the disk has finite thickness (we discuss this in \sref{sec:precess}). In our run with $r_{\rm soft}=0$, we use a purely Newtonian force $\vec{F}_{\rm N}$ outside the sinks, and transition to a Plummer force $\vec{F}_{\rm P}$ (softened using $r_{\rm sink}$) inside the sinks using a functional form $\vec{F} = \theta \vec{F}_{\rm P} + (1-\theta)\vec{F}_{\rm N}$, where $\theta = \left[1-(r/r_{\rm sink})^2\right]^2$ for distances $r<r_{\rm sink}$ from the black hole, $\theta=0$ otherwise. This regulates the singular behavior at the black hole location while achieving zero softening in the regions of interest (i.e.~regions outside the sink).

Lastly, we perform a set of ``decretion'' runs, where the binary is initialized in near-vacuum, and the sinks are replaced by sources, non-zero only within distances $r_s$ from each point mass, given by
\begin{eqnarray}
    \dot U_{\rm source} = -10^3 \Omega_K \left( \vec{U} - \vec{U}_0 \right),
\end{eqnarray}
where $\vec{U}$ are the conserved variables and $\vec{U}_0$ are their target values, given by a rigid circular rotation at speed $\sqrt{(1/2)GM/r_{\rm soft}}$, a uniform surface density $0.1 M/a^2$, and a chosen uniform Mach number. This source term strongly drives $\vec{U}$ to $\vec{U}_0$ inside the source. A circumbinary decretion disk is prevented from forming by allowing ejected material to flow off the grid. Our suite of simulations is summarized in Table~\ref{tab:simsuite}.

\subsection{Minidisk diagnostics} \label{sec:diagnostics}
To quantify minidisks individually, we integrate hydrodynamic quantities over the spatial region within a distance of $0.5 a$ from their host black hole. These diagnostics are the minidisk mass, and the center-of-mass (COM) vector measured relative to the host black hole's location. Visual inspection confirms that the COM vector points in the direction of the farthest edge of the minidisk.

\begin{figure} 
\centering
\includegraphics[width=0.47\textwidth]{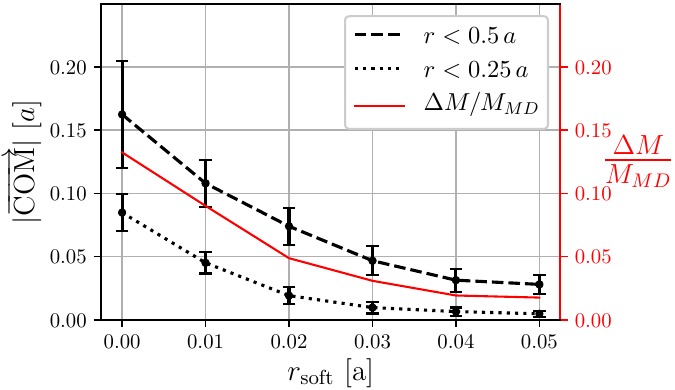}
\caption{The minidisk COM amplitude as a function of $r_{\rm soft}$, for runs using a $\Gamma$-law EOS, labeled S0 -- S5 in Table~\ref{tab:simsuite}. Error bars indicate the standard deviation of the COM amplitude over time. Two versions of the COM are shown, calculated over points within a distance of $0.5 a$ and $0.25 a$ from the host black hole. The red curve shows the mass $\Delta M$ exchanged per minidisk collision, relative to the characteristic minidisk mass $M_{MD}$ on the right axis.}
\label{fig:COMvsrsoft}
\end{figure}

\begin{figure} 
\centering
\includegraphics[width=0.47\textwidth]{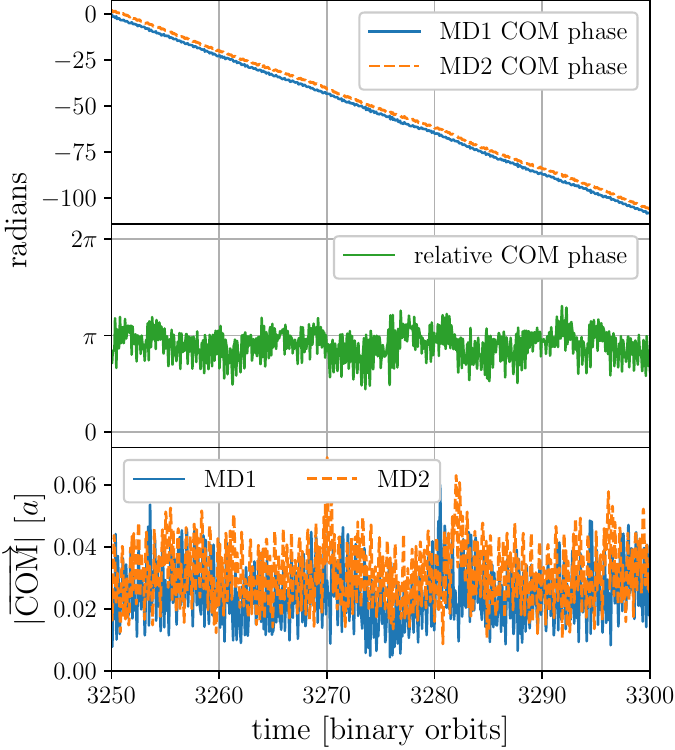}
\caption{Time series of the minidisk COM vectors for the run labeled H1 in Table~\ref{tab:simsuite}. The top panel shows the steady retrograde precession of the eccentric minidisks. The middle panel shows that the minidisks have consistently opposite orientations. The bottom panel shows the COM amplitudes of both disks.}
\label{fig:MDCOM}
\end{figure}

Finite minidisk eccentricity is found to be indicated by persistent non-zero COM amplitude over tens of binary orbits. A comparison between integrating over distances of $0.5 a$ and $0.25 a$ from black holes is provided in \fref{fig:COMvsrsoft}, and indicates that trends are robust to the size of the integration region. Crucially, the coherence of minidisk eccentricity shows up as an orderly precession, and this is indicated by a steady linear trend in the COM phase over tens of binary orbits. A tell-tale sign of a lack of persistent eccentricity is a jagged COM phase over time.
This usually indicates that the COM vector is reflecting smaller scale or more transient features in the minidisks, rather than the coherent lopsidedness characteristic of the eccentric minidisk instability. A prototypical case of steady, coherent minidisk eccentricity is shown in \fref{fig:MDCOM} (top and bottom panels).

We use two diagnostics to characterize the relationship between minidisks: relative orientation and mass flux. Their relative orientation is quantified by their relative COM phases. A prototypical case of a steady relative orientation of $\pi$ radians is shown in \fref{fig:MDCOM} (middle panel). The mass-trading between minidisks $\dot{M}_{\rm trade}(t)$ is quantified by the root-mean-square (RMS) flux of mass across a line of length $a$ through the origin, orthogonally bisecting the black hole separation.

%
%
\subsection{Solution scheme} \label{sec:numerics}
We use the \texttt{Sailfish} code, which is same code we used in W22, and we refer the reader to that work for most numerical details (\texttt{Sailfish} is a GPU implementation of \texttt{Mara3} which was used in \citealt{Tiede2020}, \citealt{Zrake:2021:eccentric} and \citealt{Tiede2022}). A summary of parameters for our suite of runs is provided in Table~\ref{tab:simsuite}. The computational domain usually extends to $12\,a$; exceptions are the high-resolution runs labeled H1$-$H4 (which extend to $15\,a$), the very high-resolution zoom-in run VH3 (which is evolved for a short time and whose grid extends to $7.5\,a$), and the decretion runs D0$-$D1 (which extend to $1.75\,a$ because there is no circumbinary disk). We use Courant-Friedrichs-Lewy numbers in the range of $0.01-0.1$. Radiative cooling requires delicate numerical treatment \citep[we follow the method of][]{Ryan+2017}. Motivated by studies which have a coordinate singularity or inner boundary between the minidisks \citep[e.g.][]{dAscoli+2018, bowen_lump, combi+2021}, we assess the effect such an obstruction by placing a third sink at the origin of the binary system of radius $0.05\,a$, in addition to the two orbiting sinks representing black holes. Note our Cartesian grid has no singular behavior at the origin.

\begin{figure} 
\centering
\includegraphics[width=0.47\textwidth]{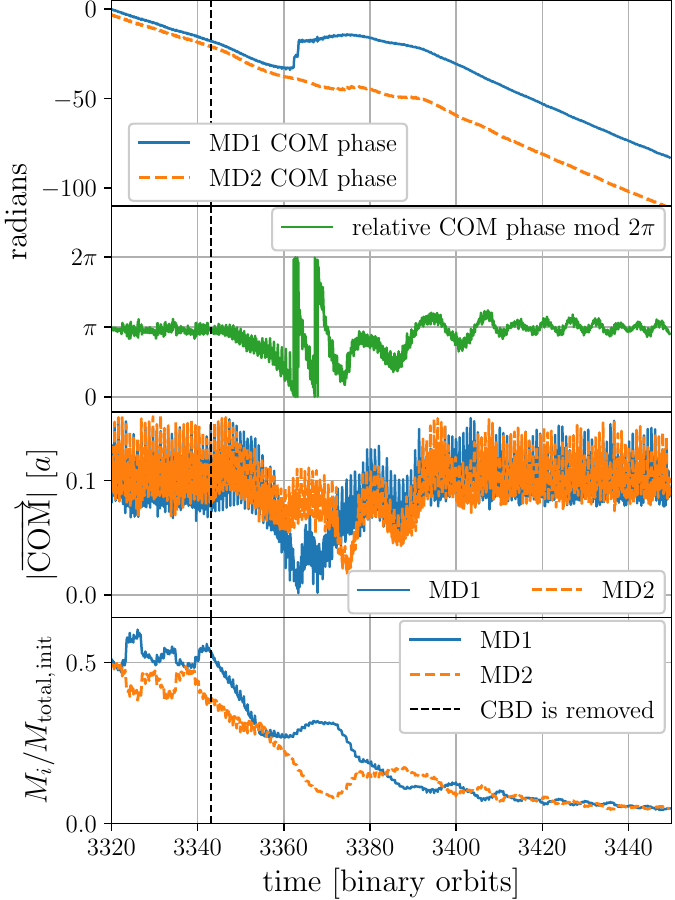}
\caption{Time series of the COM vectors and minidisk masses in a run (H2 in Table~\ref{tab:simsuite}) where the CBD is abruptly removed (vertical dashed line). Although the minidisks are steadily depleted (bottom panel), the retrograde precession (top panel), anti-alignment (top-middle), and COM amplitude (bottom-middle) eventually restore the same qualitative behavior as when the CBD was present (\fref{fig:MDCOM}).}
\label{fig:MDCOMnoCBD}
\end{figure}

%
%
\section{Results} \label{sec:results}
\begin{figure}
\begin{interactive}{animation}{anifig_noCBD.mp4}
\centering
\includegraphics[width=0.3\textwidth]{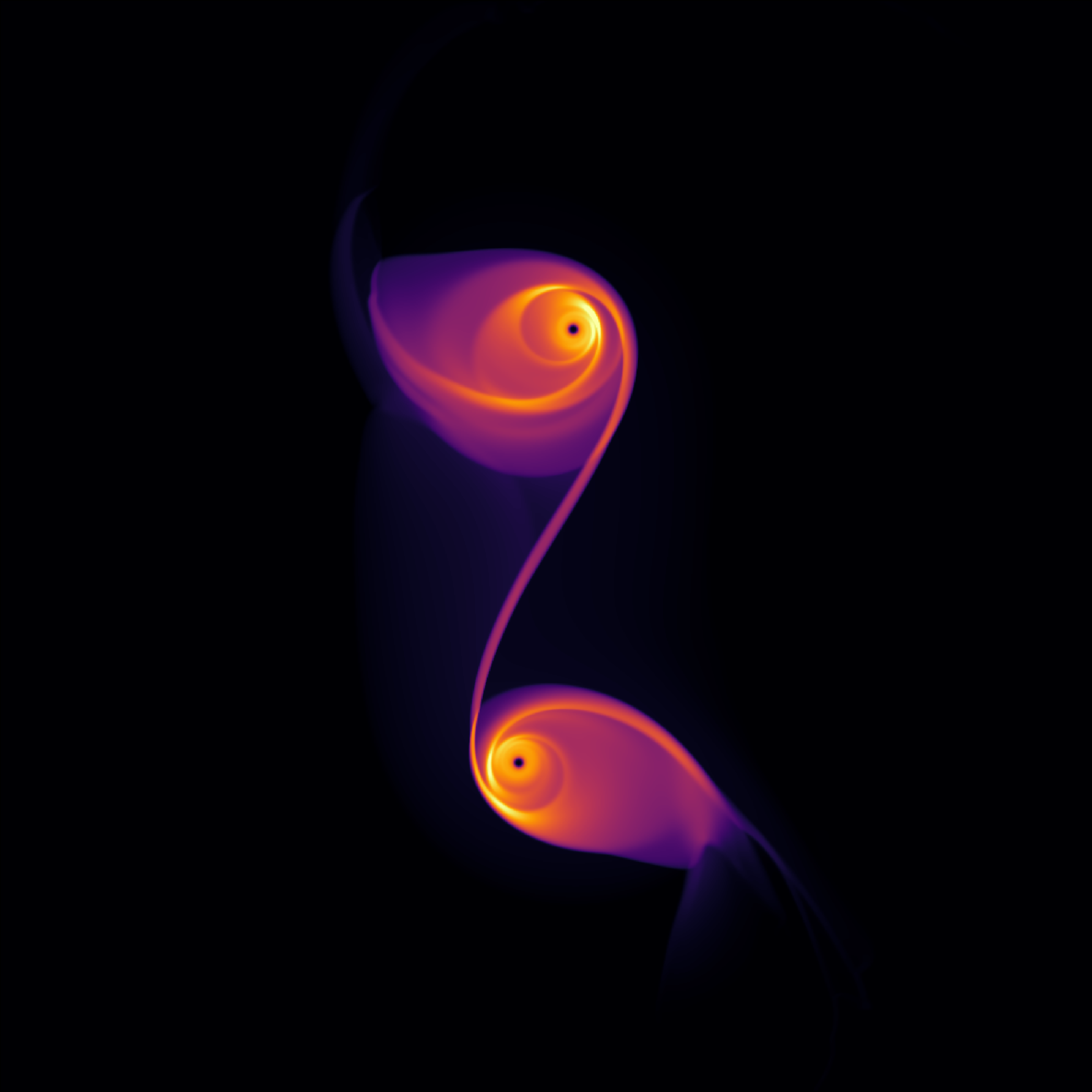}
\end{interactive}
\caption{An animated version of this figure is available in the \href{https://iopscience.iop.org/article/10.3847/1538-4357/ad1a17}{HTML version of the final article} (also available with higher quality at \url{https://youtu.be/9pltm6oOHhE}). A single frame from the animation is shown here. The animation is a 3 minute and 39 second video showing the evolution of $\sqrt{\Sigma}$ in the CBD depletion experiment from the moment the CBD is removed until about 55 binary orbits later (run H2 in Table~\ref{tab:simsuite}). The structure of the minidisks suffers a transient disruption which reestablishes over time via mass trading (see \fref{fig:MDCOMnoCBD}).}
\label{anifig:noCBD}
\end{figure}

In \fref{fig:fig1} we show snapshots of the surface density $\Sigma$ (raised to the power of $1/2$ to improve visual contrast) from two of our high-resolution runs, focusing on the minidisks. Both runs use self-consistent thermal cooling with a nominal Mach number of $\mathcal{M} \sim 11$ and $\alpha$-viscosity with $\alpha = 0.1$. The runs shown in the left and right panels of \fref{fig:fig1} are models VH3 and H2 respectively (see Table~\ref{tab:simsuite}). VH3 is a zoomed-in version of model H3, with double resolution ($\Delta x = 0.00125\,a$). The H2 and H3 models have the same parameters, except that the one on the right (model H2) has had the CBD removed, to demonstrate that the minidisks develop eccentricity even if they do not interact with gas infall from the environment. In both cases, the minidisk eccentricity is persistent, i.e.~the images in \fref{fig:fig1} are a good representation of how the disks would look at a randomly selected time in a well-evolved simulation. \fref{fig:fig1} also reveals that the minidisks settle into a configuration with their apsides oriented $180^\circ$ away from each other. In the sections below we examine these effects in detail.

\subsection{Infall from the CBD is not required to drive minidisk eccentricity}\label{sec:CBDinfall}

\begin{figure}
\begin{interactive}{animation}{anifig_dec.mp4}
\centering
\includegraphics[width=0.3\textwidth]{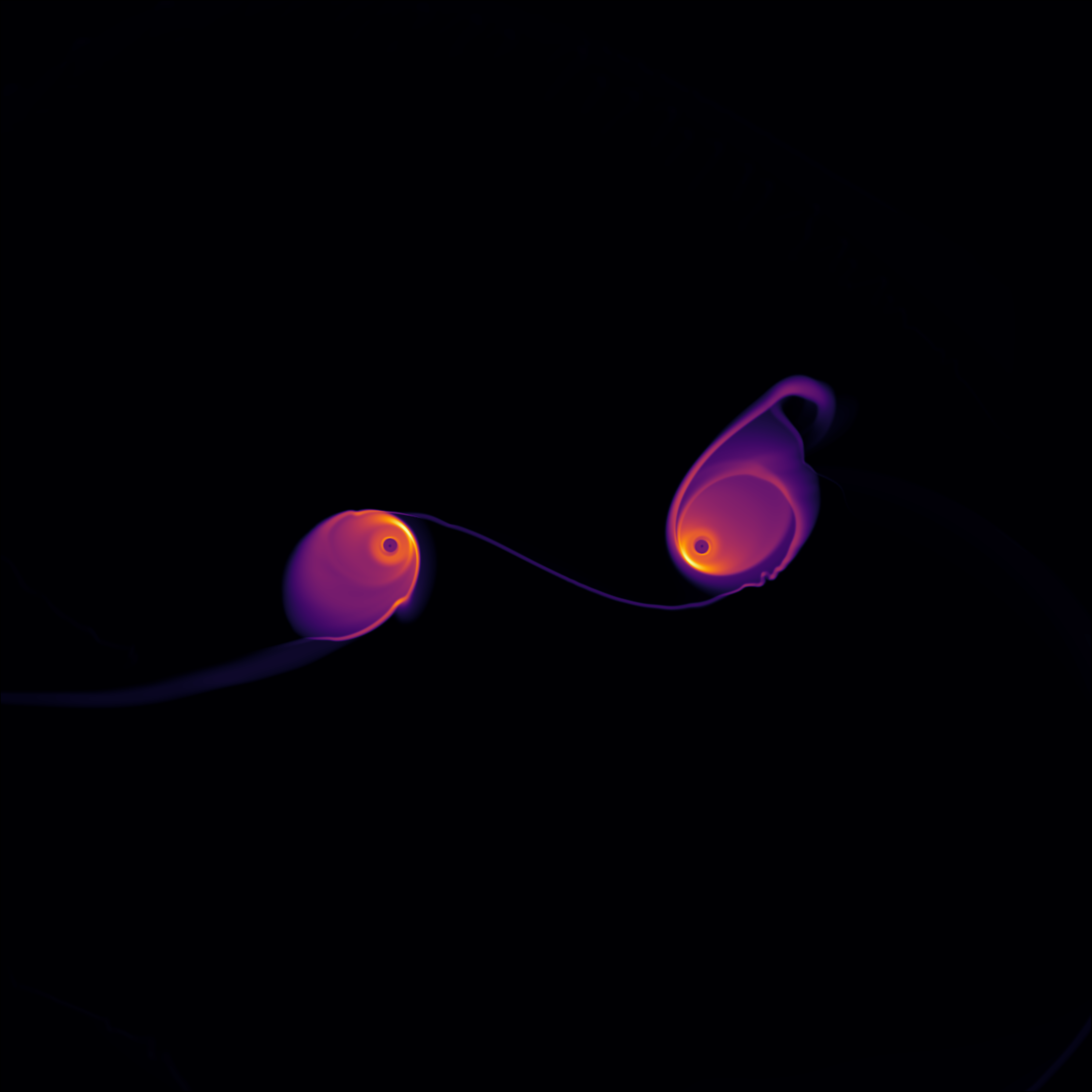}
\end{interactive}
\caption{An animated version of this figure is available in the \href{https://iopscience.iop.org/article/10.3847/1538-4357/ad1a17}{HTML version of the final article} (also available with higher quality at \url{https://youtu.be/om15kZRhC18}). A single frame from the animation is shown here. The animation is a 3 minute and 20 second video showing the evolution of $\sqrt{\Sigma}$ for all 50 binary orbits in the decretion experiment (run D1 in Table~\ref{tab:simsuite}). The minidisks form from the inside out, fill their Roche lobes, and begin a synchronized mass trade, resulting in eccentricity growth (see \fref{fig:MDCOMdecretion}).}
\label{anifig:dec}
\end{figure}

\begin{figure} 
\centering
\includegraphics[width=0.47\textwidth]{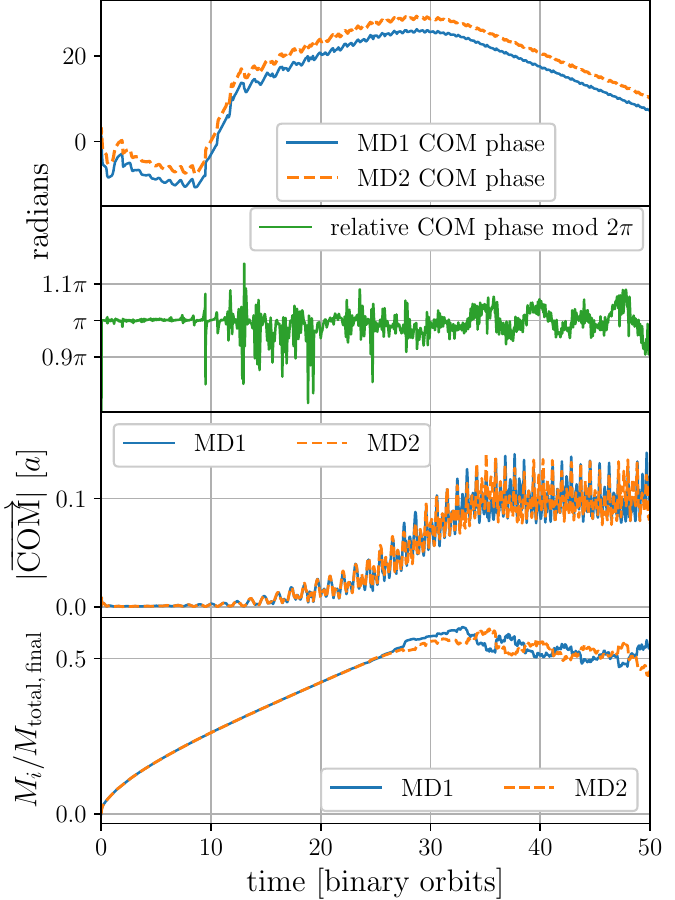}
\caption{Time series of the COM vectors and minidisk masses in a run where the CBD is removed and mass is supplied from the component locations (``decretion'' run, D1 in Table~\ref{tab:simsuite}). The behavior is largely the same as the CBD removal experiment shown in \fref{fig:MDCOMnoCBD}. Note the top-middle panel is zoomed in on $\pi$, compared to \fref{fig:MDCOMnoCBD}. The relative orientation of the minidisks begins at $\pi$ due to the mirror symmetry of the system.}
\label{fig:MDCOMdecretion}
\end{figure}

In order to determine whether the minidisk eccentricity is driven by gas infall from the CBD, we restarted a run from a well-developed state, with the CBD subtracted and replaced with a near-vacuum (model H2). After the restart, the minidisks continue to evolve, but can no longer acquire gas from the environment. The right panel of \fref{fig:fig1} shows that the eventual relaxed state of the minidisks is again eccentric, and still has the disk apsides anti-parallel to one another. \fref{fig:MDCOMnoCBD} shows the time series of the minidisk diagnostics following the depletion. Without feeding from the CBD, the mass in the minidisks diminishes over time as they accrete into the sinks (4th panel). The minidisk eccentricity (indicated by COM amplitude, 3rd panel), opposing orientation (2nd panel), and precession (1st panel) undergo a short disruption at roughly 20 binary orbits after the CBD is depleted, but minidisk eccentricity then restores over the subsequent 40 binary orbits. The maintenance of eccentric minidisks in the absence of gas infall from the CBD indicates that eccentricity is being injected by interactions between the minidisks. A video showing this run is provided in \fref{anifig:noCBD}. 

We have also confirmed that the eccentric minidisks can be established if mass is supplied from the sinks in a ``decretion'' run (model D1). For this model, a circular equal-mass binary was initialized in near-vacuum, with mass being steadily added to the system from the sinks (rather than subtracted, see \sref{sec:models}). Animations of the decretion run (see \fref{anifig:dec})
illustrate how the eccentric minidisks are established. First, gas flows out from the particle positions and forms minidisks around the binary components. Then as the minidisks grow in size and overflow their Roche lobes, they ``collide'' and exchange mass across the inner Lagrange point. After the mass-trading event, the minidisks recede inside their Roche lobes, but develop a small amount of eccentricity. Subsequently, the minidisks collide preferentially at their ``long end,'' leading to further eccentricity injection and then stronger collisions. \fref{fig:MDCOMdecretion} shows the time series of minidisk diagnostics in the decretion experiment, and indicates that the eccentric minidisks, retrograde precession, and anti-parallel orientations become fully established.

\begin{figure} 
\centering
\includegraphics[width=0.47\textwidth]{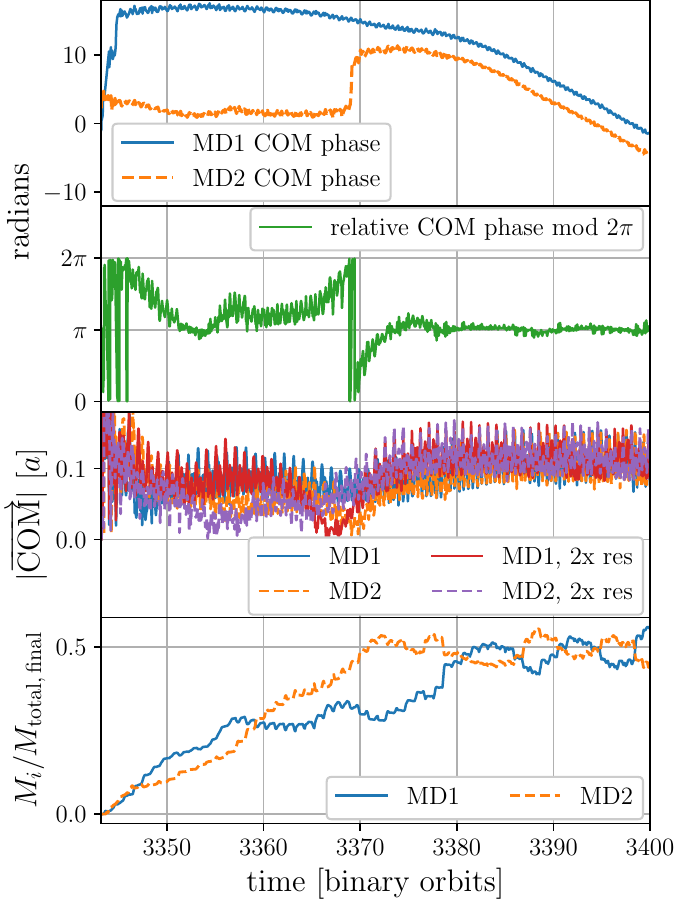}
\caption{Similar to \fref{fig:MDCOM}, but for the minidisk refilling experiment run with a smaller softening length of $r_{\rm soft}=0.01\,a$ (run H3 in Table~\ref{tab:simsuite}). The VH3 run with double the resolution of H3 is also shown in the middle-bottom panel.}
\label{fig:MDCOMnoMDs}
\end{figure}

\begin{figure}
\begin{interactive}{animation}{anifig_noMDs.mp4}
\centering
\includegraphics[width=0.3\textwidth]{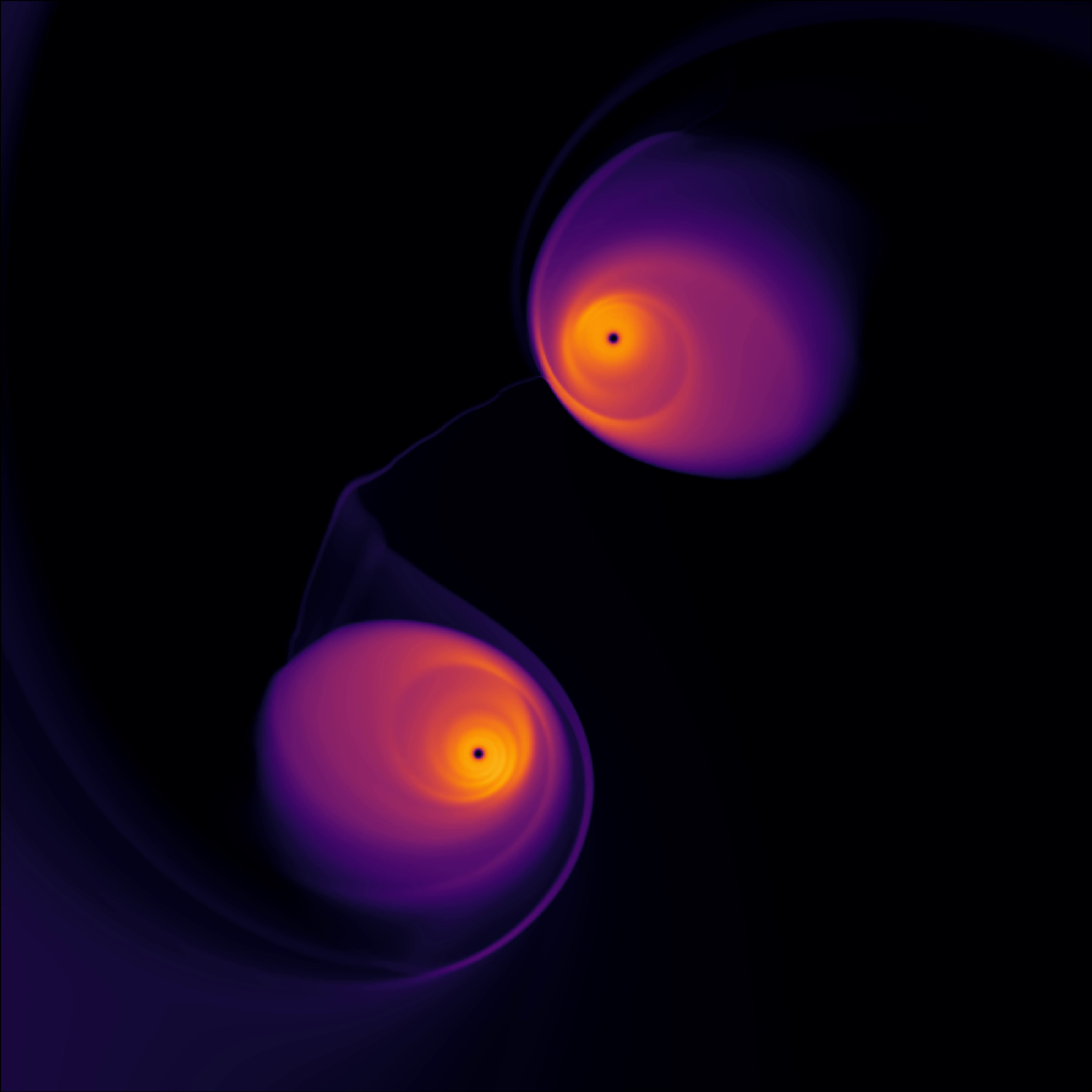}
\end{interactive}
\caption{An animated version of this figure is available in the \href{https://iopscience.iop.org/article/10.3847/1538-4357/ad1a17}{HTML version of the final article} (also available with higher quality at \url{https://youtu.be/GCh7yW-QuY8}). A single frame from the animation is shown here. The animation is a 3 minute and 46 second video showing the evolution of $\sqrt{\Sigma}$ over roughly 60 binary orbits in the minidisk refilling experiment (run H3 in Table~\ref{tab:simsuite}). Mass is drawn from the CBD to reform the minidisks, which fill their Roche lobes and begin a synchronized mass trade, resulting in eccentricity growth (see \fref{fig:MDCOMnoMDs}).}
\label{anifig:noMDs}
\end{figure}

\fref{fig:MDCOMnoMDs} shows the results of one final experiment, in which the minidisks are subtracted but the CBD gas is retained (model H3). The results are similar to the decretion run: the minidisks refill, this time from gas infalling from the CBD, and over the course of about 30 orbits they settle into the characteristic eccentric, anti-aligned configuration. In the 3rd panel we also show the model VH3, which is a zoomed-in version of H3 with double resolution ($\Delta x = 0.00125\,a$), and which shows that the minidisk eccentricity settles to the same level. A video of this run is provided in \fref{anifig:noMDs}.

\pagebreak
\subsection{Role of the minidisk-minidisk interaction} \label{sec:interaction}

\begin{figure} 
\centering
\includegraphics[width=0.47\textwidth]
{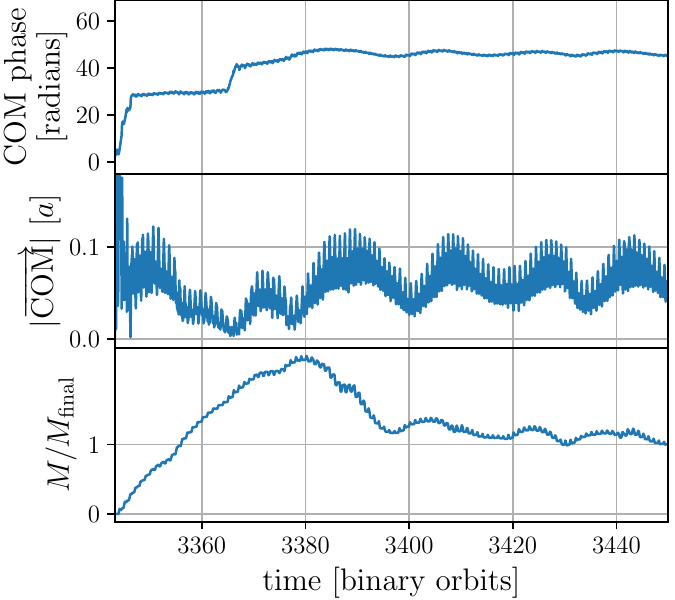}
\caption{Similar to \fref{fig:MDCOMnoMDs}, but with one black hole replaced with an absorber of radius $0.45 a$ (run H4 in Table~\ref{tab:simsuite}). Without a companion, the minidisk develops some eccentricity, but the magnitude is smaller and more variable in time.}
\label{fig:MDCOMnoMDs1absbig}
\end{figure}

\begin{figure}
\begin{interactive}{animation}{anifig_noMDs_1absbig.mp4}
\centering
\includegraphics[width=0.3\textwidth]{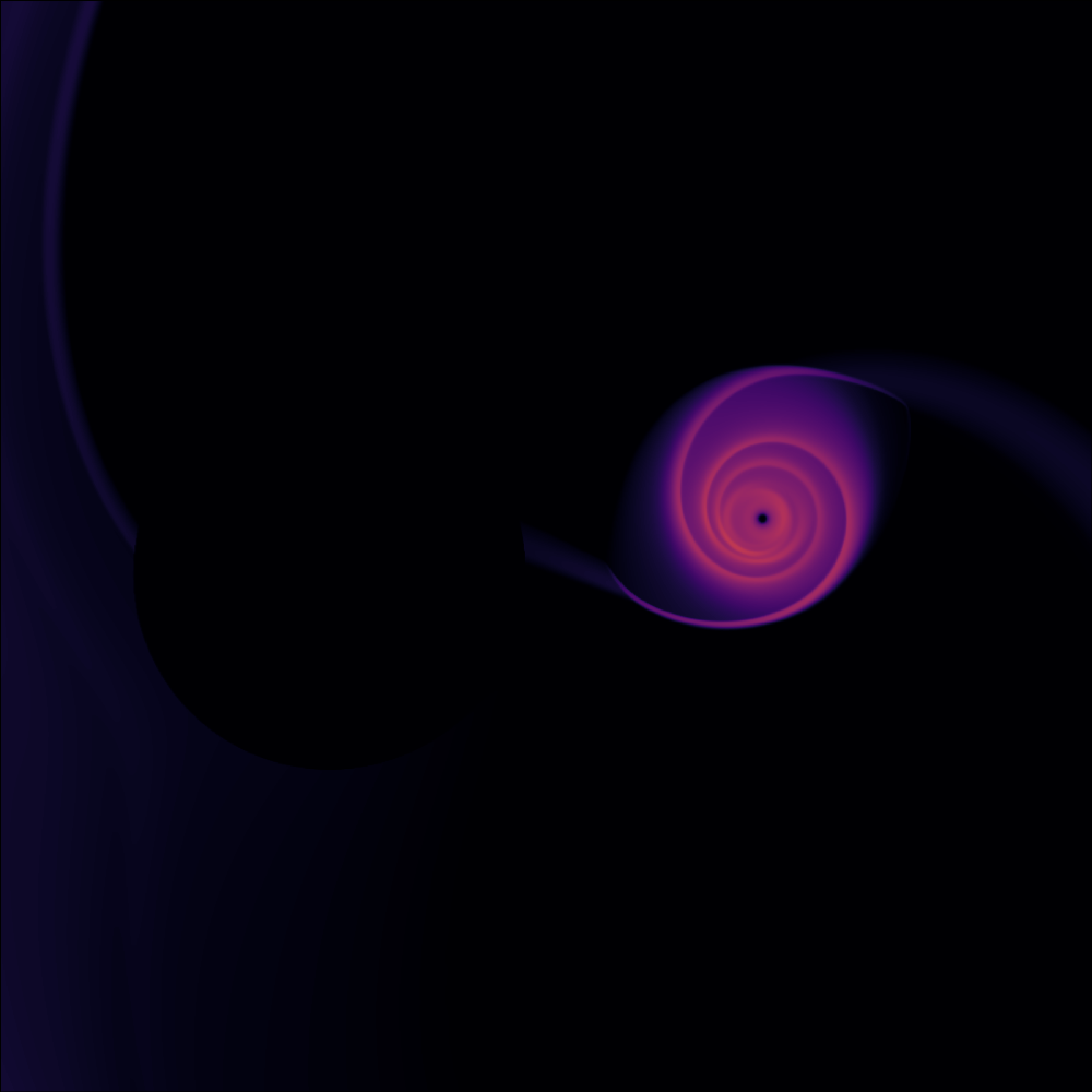}
\end{interactive}
\caption{An animated version of this figure is available in the \href{https://iopscience.iop.org/article/10.3847/1538-4357/ad1a17}{HTML version of the final article} (also available with higher quality at \url{https://youtu.be/Th13XvxKsxA}). A single frame from the animation is shown here. The animation is a 3 minute and 46 second video showing the evolution of $\sqrt{\Sigma}$ over roughly 60 binary orbits in the single minidisk refilling experiment (run H4 in Table~\ref{tab:simsuite}). Mass is drawn from the CBD to reform the lone minidisk, which fills its Roche lobe. Without mass trading from a companion minidisk, it does not develop a persistent value of eccentricity (see \fref{fig:MDCOMnoMDs1absbig}).}
\label{anifig:noMDS_1absbig}
\end{figure}

The visual impression given by animations of the decretion and minidisk refilling experiments (models D1 and H3) is that interaction \emph{between} the minidisks, and the associated mass exchanges, mediate the eccentricity growth. To see how things would be changed without the minidisk-minidisk interaction, we performed a run (model H4) where one of the minidisks is replaced by a large absorber of radius $0.45\,a$. In this configuration, one minidisk refills from the circumbinary disk, and can lose mass to the companion absorber, but does not receive stream impacts from a companion minidisk. Time series of the minidisk diagnostics in \fref{fig:MDCOMnoMDs1absbig} show that the COM amplitude of the lone minidisk (2nd panel) exhibits large oscillations around roughly $0.05a$. A video of this run is provided in \fref{anifig:noMDS_1absbig}.
In contrast, with no absorber present (model H3; \fref{fig:MDCOMnoMDs}), the COM amplitude is about $0.1a$ with relatively little variation. Also, the minidisks undergo a steady rate of retrograde precession in the ``normal'' run H3, and that precession is not seen when the absorber is present (\fref{fig:MDCOMnoMDs} top panel vs. \fref{fig:MDCOMnoMDs1absbig} top panel). These observations indicate that some eccentricity must be injected by gas infall to the minidisks, but not persistently enough to account for the minidisks observed in the ``normal'' run H3. The persistent eccentricity, seen in runs that include both minidisks, indicates the minidisk-minidisk coupling is a likely cause of the directionally coherent eccentricity injection.

\subsection{Periodic mass trade and apsidal precession} \label{sec:masstrade}
The eccentric minidisks collide periodically and exchange mass. This effect is shown quantitatively in the top panel of \fref{fig:mdot_vs_mtrade}, where we plot the RMS mass flux across the midline between the binary components (the midline rotates at the binary orbital frequency). The spikes in the RMS mass flux correspond to a rate of mass exchange that exceeds the average mass flow to the binary by factors of $10 - 20$. The mass transferred per collision exceeds 20\% of the disk mass when the instability is most aggressive (\fref{fig:COMvsrsoft}), so the mass-trading events are dynamically significant. The pulses are very regular, and the interval is on the order of the binary orbital period.

The frequency of the mass trading events is accurately predicted by the beat frequency $f_{\rm bin} - f_{\rm prec}$ associated with the binary orbital frequency $f_{\rm bin}$ and the apsidal precession frequency $f_{\rm prec}$ of the minidisks. This is confirmed in \fref{fig:rsoftlim}, which shows periodograms of the RMS mass flux between minidisks for runs S0-S5 (same runs as the first row of \fref{fig:MD_tiles}).

Apsidal precession of eccentric disks in binary systems is generally governed by a combination of pressure gradients, viscous stresses, and the tidal field of the companion \citep[see e.g.][]{Goodchild+2006, kley+2008}. The precession rate seen in our simulations is additionally found to be sensitive to the gravitational softening length, $r_{\rm soft}$. For example, the run shown in the left-most panel of the top row of \fref{fig:MD_tiles} has a precession rate that is consistent with zero, and that run (model S0) also has a zero gravitational softening length. We also performed a decretion run with zero softening (model D0 in Table~\ref{tab:simsuite}) where the initial condition is a low-density atmosphere and gas is injected from the component locations, and with a small viscosity $\alpha=0.001$. This experiment reveals that kinematic shear viscosity tends to drive retrograde disk precession; in contrast to model S0 (which has $\alpha=0.1$), in model D0, which has much lower viscosity, we found the minidisk precession becomes prograde with a period of $\sim 47$ binary orbits. A video showing this run is provided in \fref{anifig:dec_a0001}.
When gravitational softening is zero and the viscosity is negligible, the precession rate can still be positive or negative, and is then determined by a competition between tidal interaction with the companion (which drives prograde precession) and pressure gradients, which generally drive retrograde precession provided that radial derivatives of pressure are not too positive; \cite{Goodchild+2006}.

\subsection{Dependence on gravitational softening} \label{sec:depsoft}
To determine the necessary conditions for the instability to operate, we have systematically ``switched off'' different pieces of physics. A summary of the visual results is presented in the panels of \fref{fig:MD_tiles}. The top row shows a sequence of representative images (again, color showing $\Sigma^{1/2}$) from runs where the gravitational softening length $r_{\rm soft}$ is increased from $0.0$ to $0.05a$. It is visually evident that the instability gets weaker with larger $r_{\rm soft}$, and becomes too weak to see when $r_{\rm soft} \gtrsim 0.03a$. This trend is corroborated in \fref{fig:COMvsrsoft}, where we plot the distance of the minidisk COM from the respective binary component (as described in \ref{sec:diagnostics}) as a function of $r_{\rm soft}$. Although the minidisk eccentricity is not visually obvious for $r_{\rm soft}\gtrsim 0.03a$, precession is nonetheless quantifiable (see e.g.~the high-resolution model H1 with $r_{\rm soft}=0.04\,a$ in \fref{fig:MDCOM}), and \fref{fig:rsoftlim} shows the cadence of minidisk mass exchange is still well-predicted by $f_{\rm bin} - f_{\rm prec}$.

\subsection{Dependence on the orbital Mach number} \label{sec:depmach}
The second row of images in \fref{fig:MD_tiles} shows representative minidisk morphologies for a range of nominal Mach numbers in the range $7 - 25$, and significant and persistent minidisk eccentricity is seen for all cases. The top panel of \fref{fig:COMvsnuMach} shows the minidisk COM diagnostic as a function of the Mach number, and confirms that there is no clear dependence of the minidisk eccentricity on the disk temperature in the range we have simulated here.

\subsection{Dependence on gas viscosity and suppression by target temperature profiles}
\label{sec:depvisc}
The third row of \fref{fig:MD_tiles} shows how the minidisk morphology depends on the gas viscosity, and reveals that high enough viscosity, $\nu \gtrsim 10^{-4}$ reliably suppresses the instability, resulting in roughly circular minidisks. This is also corroborated in the bottom panel of \fref{fig:COMvsnuMach}. The fourth row of \fref{fig:MD_tiles} (other than the right-most image) shows that the instability is significantly suppressed by the use of target temperature profiles. The degree of suppression is not markedly affected by the rate of driving towards the target temperature profile (compare 2nd and 3rd panels). Crucially, use of the isothermal equation of state (4th row, 4th panel), which is widely used in studies of binary accretion, strongly suppresses minidisk eccentricity in circumbinary accretion. However, in \sref{sec:convergence} we show that suppression by the isothermal equation of state can be overcome in some cases.

\begin{figure} 
\centering
\includegraphics[width=0.47\textwidth]{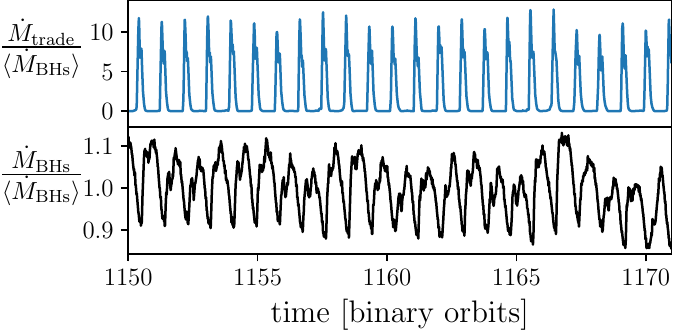}
\caption{Top: Time series of the inter-minidisk mass trading rate $\dot{M}_{\rm trade}$. Bottom: Time series of the instantaneous mass accretion rate $\dot{M}_{\rm BHs}$ to the black holes. Both signals are normalized by the time-averaged mass accretion rate $\langle \dot{M}_{\rm BHs} \rangle$, and are computed from our run S1.}
\label{fig:mdot_vs_mtrade}
\end{figure}

\subsection{Effect of a ``hole'' near the origin} \label{sec:dephole}
Many studies of binary accretion use a grid code with cylindrical polar coordinates, and such geometries could induce anomolous flow patterns in the vicinity of the coordinate origin. Given that mass transferred between the minidisks generally passes through the origin, we found it germane to examine how a source of systematic numerical error, such as arising from a coordinate singularity or inner boundary, might affect how the instability behaves.
We modeled the source of error using a ``hole'' placed at the origin (runs labeled HOLE and S2 in Table~\ref{tab:simsuite}), which is included as a third sink term as described in \sref{sec:numerics}.

The minidisk morphology when the hole is present is shown at the bottom right panel of \fref{fig:MD_tiles}. The time series of the minidisk COM, shown in \fref{fig:hole}, shows that the hole diminishes the average minidisk eccentricity by about half, and also reveals a large amplitude, slow oscillation about the mean eccentricity. This suggests the instability could be mischaracterized in simulations that use a cylindrical polar coordinate grid, unless particular care is taken to avoid numerical errors near the origin.

This experiment also yielded serendipitous insight into the dynamics of the eccentricity driving. The slow oscillation of the minidisk COM is seen in both disks, but these oscillations are $180^\circ$ out of phase with one another; one disk gets more eccentric while the other circularizes. The oscillation period for the run shown in \fref{fig:hole} is roughly 9 orbits, which is also the minidisk apsidal precession period for that run. We now understand that the radializing minidisk is hogging the gas falling in from the CBD, and that the circularizing minidisk is relatively starved. The explanation for this may be as follows: (a) the CBD cavity is eccentric, (b) the minidisks are eccentric, and (c) one of the disks extends more in the direction of the near side of the cavity wall, thereby receiving more of the infalling gas. These conditions apply too when no hole is present, but then gas flows relatively unimpeded from the disk which catches more of the infall to the one which catches less, and both disks remain fed. By inhibiting the mass and momentum transfer between disks, the hole leads to the starvation of one disk at a time, and also allows that disk to circularize. This is a further indication that the minidisk-minidisk interaction (\sref{sec:interaction}) is instrumental in driving the instability.

\begin{figure} 
\centering
\includegraphics[width=0.47\textwidth]{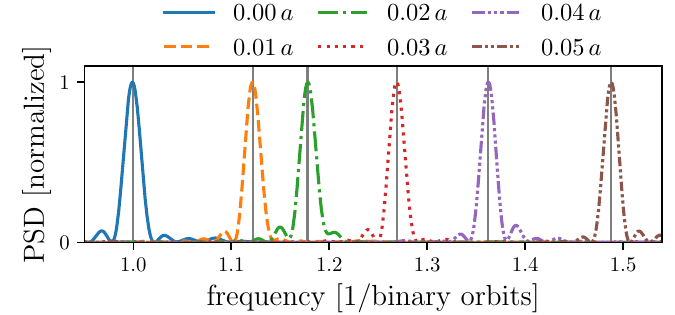}
\caption{Normalized PSDs of the RMS mass flux between minidisks for different values of $r_{\rm soft}$. Data is from the fiducial runs with $\Gamma$-law EOS shown in \fref{fig:COMvsrsoft}. Vertical solid lines indicate the ``eccentric minidisk beat frequency'' $f_{\rm bin} - f_{\rm prec}$, computed from the minidisk COM phase evolution, accurately predicting the peak frequencies.}
\label{fig:rsoftlim}
\end{figure}

\begin{figure*}[ht]
\centering
\includegraphics[width=1\textwidth]{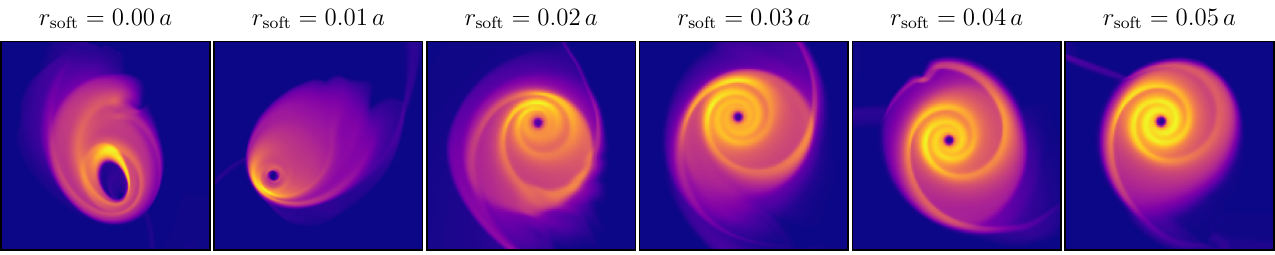}\\
\includegraphics[width=1\textwidth]{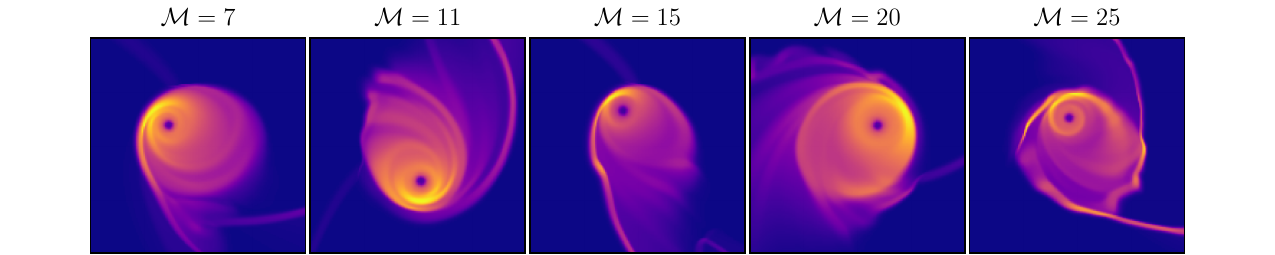}\\
\includegraphics[width=1\textwidth]{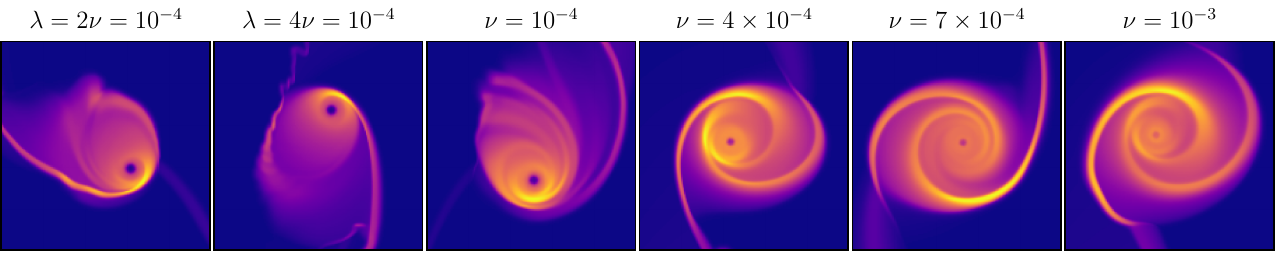}\\
\includegraphics[width=1\textwidth]{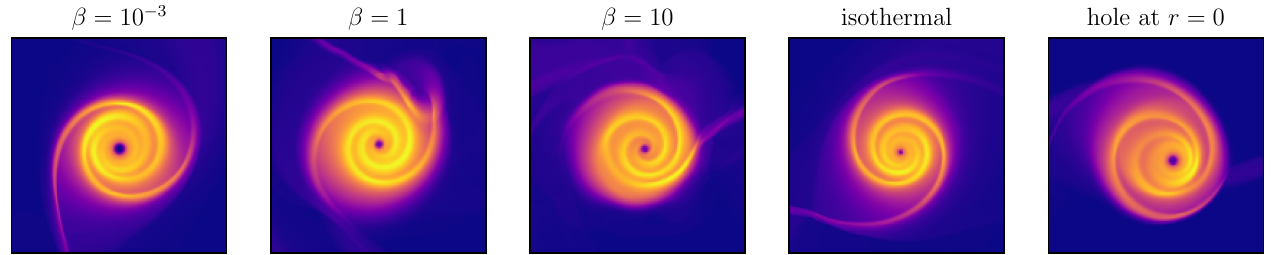}
\caption{Representative minidisk snapshots, displaying $\sqrt{\Sigma}$ on independent color scales. Each tile spans a square region with side lengths $0.9\, a$. Kinematic shear and bulk viscosities ($\nu$ and $\lambda$) are given in units of $\sqrt{GMa}$. Runs listed in Table~\ref{tab:simsuite} correspond to this figure as follows. First row: S0-S5. Second row: M7-M25. Third row: BV2, BV4, V1-V10. Fourth row: $\beta$m3, $\beta$0, $\beta$1, ISO, HOLE.} \label{fig:MD_tiles}
\end{figure*}

\begin{figure}
\begin{interactive}{animation}{anifig_dec_a0001.mp4}
\centering
\includegraphics[width=0.3\textwidth]{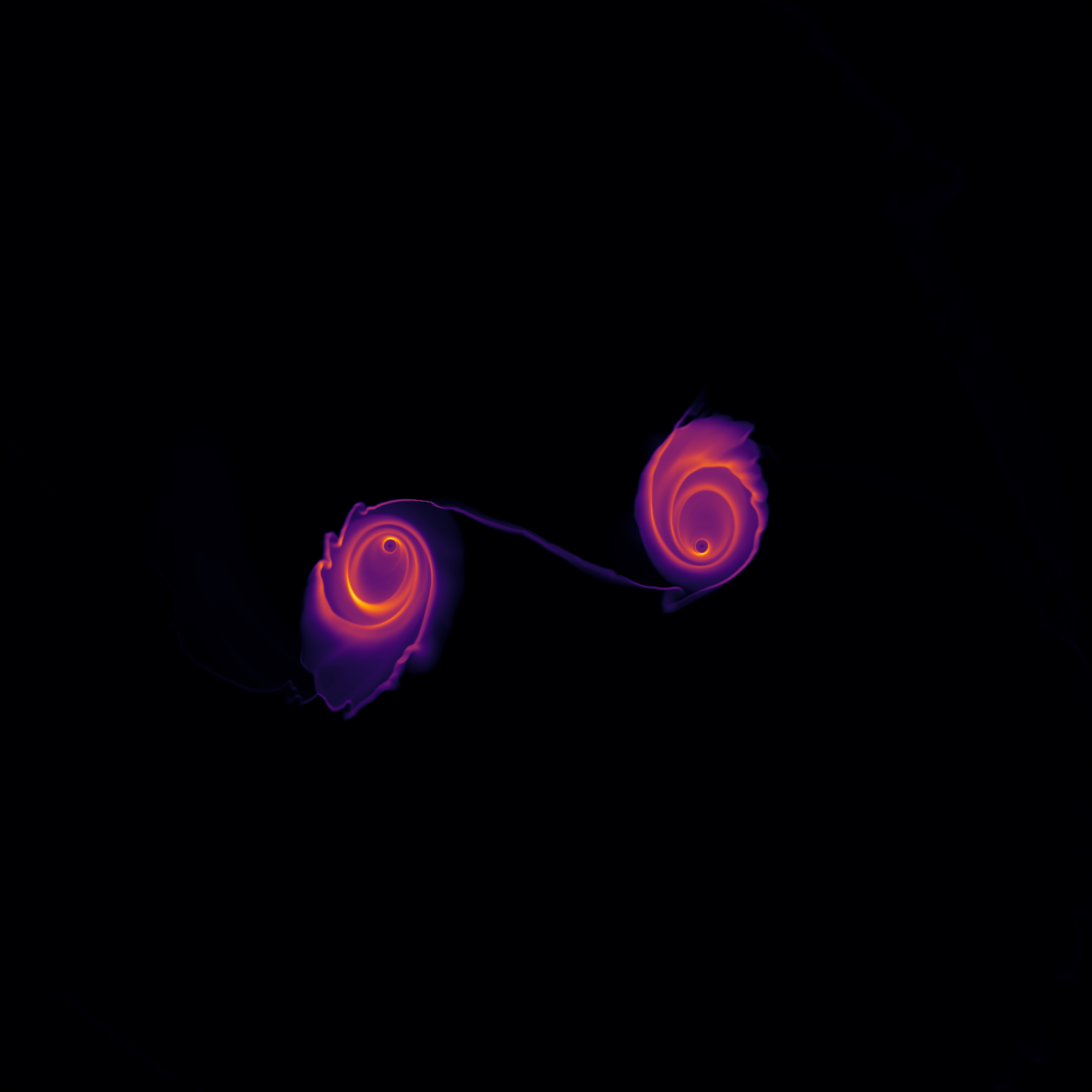}
\end{interactive}
\caption{An animated version of this figure is available in the \href{https://iopscience.iop.org/article/10.3847/1538-4357/ad1a17}{HTML version of the final article} (also available with higher quality at \url{https://youtu.be/rFEvJFRePDA}). A single frame from the animation is shown here. The animation is a 3 minute video showing the evolution of $\sqrt{\Sigma}$ over 50 binary orbits in the low-viscosity ($\alpha=0.001$), zero-softening decretion experiment (run D0 in Table~\ref{tab:simsuite}). The minidisks form from the inside out, fill their Roche lobes, and begin a synchronized mass trade, resulting in eccentricity growth, but in this case the minidisks precess slowly in a prograde sense.}
\label{anifig:dec_a0001}
\end{figure}

\subsection{Numerical convergence of eccentricity growth rates} \label{sec:convergence}
We performed a resolution study of the exponential growth rate of the minidisk COM amplitude using decretion experiments. We used the decretion scenario in order to remove the complicating effects of the mass infall from the CBD. Indeed, this scenario most cleanly isolates the role of the minidisk-minidisk interaction in driving the instability. 
When the CBD is removed, the instability occurs even with an isothermal EOS.\footnote{Note that we have not observed robust development of the instability in any runs that both use a target temperature profile and include the CBD. The CBD introduces an asymmetry in mass feeding to the minidisks by feeding them one at a time, and the degree of its influence on the minidisk-minidisk mass exchange is evidently sensitive to the cooling model. The runs shown in \fref{fig:dec} have nearly perfect symmetry in the minidisk-minidisk mass exchange, and therefore indicate that a high degree of such symmetry plays an important role in this eccentric instability. Why CBDs suppress minidisk eccentricity when using target temperature profiles, but not otherwise, is not yet clear -- see the discussion at the end of \sref{sec:mechanism}}. 
Since isothermal runs are significantly less computationally expensive than radiatively cooled runs, we used the isothermal EOS with no CBD to illustrate the numerical convergence of the growth rate over a wider range of resolutions. \fref{fig:dec} shows the exponential growth of the minidisk COM amplitude, in runs where the grid resolution is 200, 400, 800, 1600, and 3200 zones per semi-major axis. All of these runs develop eccentric minidisks, but the lowest resolution runs displayed in each panel, 200 (400) zones per $a$ for an isothermal (radiatively cooled $\Gamma$-law) equation of state, show a spuriously large growth rate and early saturation. The growth rate is consistently measured to be approximately $0.07 f_{\rm bin}$ at each subsequent doubling of the grid resolution. Saturation occurs around a consistent value. Note that the series of convergence runs displayed in \fref{fig:dec} is not enumerated in Table~\ref{tab:simsuite}.

\subsection{Summary of key numerical findings} \label{sec:results-summary}
The results of our numerical investigation strongly point to the minidisk-minidisk interaction as a necessary and sufficient condition for the growth of persistent minidisk eccentricity. Namely, in runs which only have minidisks and their unimpeded interaction (models H2 and D1, \sref{sec:CBDinfall}), minidisk eccentricity is seen to grow and saturate at a persistent level, showing that the minidisk interaction is sufficient to generate persistent eccentricity; whereas shutting off the minidisk-minidisk interaction (model H2, \sref{sec:interaction}), or impeding their interaction (model HOLE, \sref{sec:dephole}), prevents persistent minidisk eccentricity. The minidisk eccentricity growth rate numerically converges (\sref{sec:convergence}), showing it is a physical effect. The minidisk-minidisk interaction manifests as the regular trading of mass between the minidisks across the inner Lagrange point, at roughly the orbital period of the binary. Departure of the observed mass trade interval from the orbital period is due to apsidal precession of the minidisks (\sref{sec:masstrade}). Precession can in general be prograde or retrograde, but when $r_{\rm soft} \gtrsim 0.01a$ it is always retrograde and gets faster with increased $r_{\rm soft}$ (\sref{sec:depsoft}). This effect is consistent with known retrograde precession of ballistic particles in a softened gravitational potential, which we also checked with numerical integrations of eccentric particle orbits in softened potentials.
The instability likely exists in a formal sense regardless of how thermodynamics is modeled, however it seems to be suppressed in scenarios where a CBD is present, and a target temperature profile is used, as with $\beta$-cooling or the locally isothermal EOS.

%
%
\begin{figure} 
\centering
\includegraphics[width=0.47\textwidth]{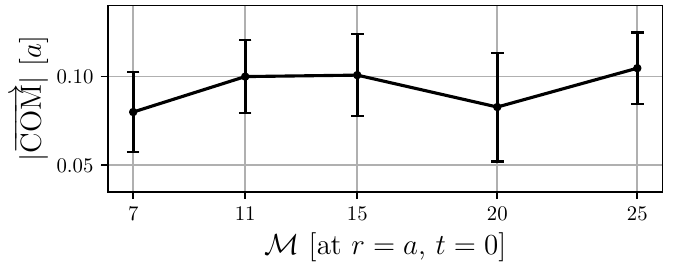}\\
\includegraphics[width=0.47\textwidth]{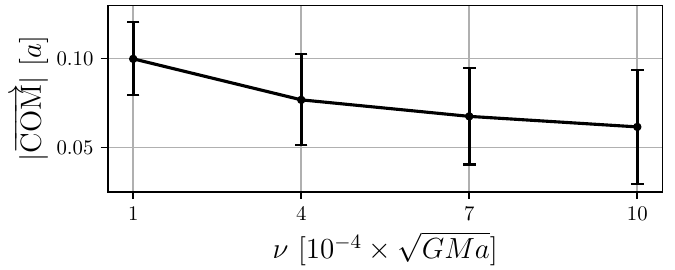}
\caption{Top: Time-averaged COM amplitude of minidisks versus Mach number $\mathcal{M}(r\!=\!a,t\!=\!0)$, for runs using a $\Gamma$-law equation of state (runs M7 -- M25 in Table~\ref{tab:simsuite}). Bottom: Time-averaged COM amplitude of minidisks versus kinematic viscosity $\nu$, for runs using a $\Gamma$-law equation of state (runs V1 -- V10 in Table~\ref{tab:simsuite}). Error bars correspond to the standard deviation of the COM amplitude.}
\label{fig:COMvsnuMach}
\end{figure}

\section{Discussion}\label{sec:discuss}

\subsection{Mechanism of the instability} \label{sec:mechanism}
Our numerical calculations in \sref{sec:results} clearly point to a mechanism for minidisk eccentricity growth: we propose that minidisk eccentricity is injected by regular impacts between the minidisks satisfying a certain resonant phase condition.
In this section, we present a heuristic picture of this whereby the outer edge of a minidisk is pictured as an eccentric ring of test particles in Keplerian orbit around a binary component, as illustrated in \fref{fig:diagram}. The particles in the ring are subject to an external forcing term $\vec{f}_e(\nu, \theta)$, which depends on the ring eccentricity $e$, and the orbital phases, $\nu$ and $\theta$, of the ring particle and of the binary orbit respectively. The zero of the binary orbital phase is chosen so that $\theta=0$ means the eccentricity vector $\vec{e}_1$ of the BH1 minidisk points horizontally to the right.

The forcing term needs to capture the dynamical effects of head-on impacts between gas parcels in opposing minidisks. Since more mass is exchanged per impact as the eccentricity grows (\fref{fig:COMvsrsoft}), the forcing amplitude must increase with $e$.
Impacts occur when $\theta \simeq 0$, and for small eccentricities they mainly affect the particles at the long ends of the minidisks around $\nu \simeq \pi$. The impact force is directed opposite the particle's velocity $\vec{v}_{\rm orb}(\nu)$, and is proportional in magnitude to its speed. A possible forcing term is then
\begin{equation} \label{eqn:ring-forcing}
\vec{f}_e(\nu, \theta) = -{\rm const} \times e(t) \times \ \delta(\theta) \ \delta(\nu - \pi) \ \vec{v}_{\rm orb}(\nu) \, ,
\end{equation}
where $\delta$ is a Dirac delta function and the constant in Eqn.~\ref{eqn:ring-forcing} is positive. 
The ring evolves ``rigidly'', in the sense that, only that part of the forcing which determines the total torque and power applied to the ring is included in the equation of motion. This also means the ring does not precess, although one could estimate the rotation rates of $\vec{e}_{1,2}$ by averaging the local rate of apsidal rotation over the particle phases $\nu$; non-elliptical distortions obviously cannot be captured in the ring approximation. Integration of $\vec r \times \vec{f}_e $ and $\vec{f}_e \cdot \vec{v}_{\rm orb}$ over $d\nu$ yields respectively the ring specific torque $\dot \ell$ and specific power $\dot {\cal E}$, and in turn, an expression for $\dot e(t)$ via $e = \sqrt{1 + 2 {\cal E} \ell^2 / (G M)^2}$.

\begin{figure} 
\centering
\includegraphics[width=0.47\textwidth]{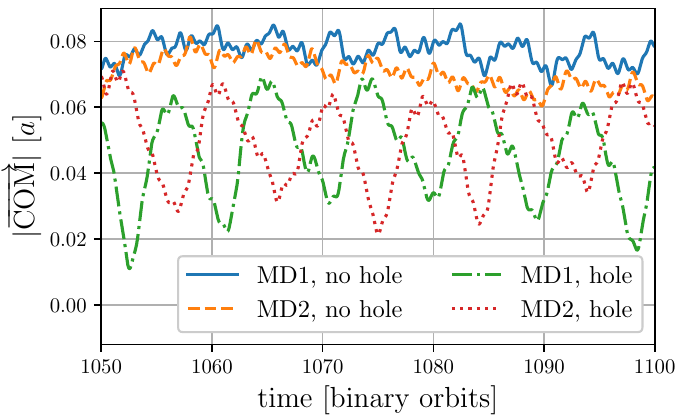}
\caption{Time series of the COM amplitude of minidisks, for runs with and without a ``hole'' at $r<0.05 a$ (runs labeled HOLE and S2 in Table~\ref{tab:simsuite}).}
\label{fig:hole}
\end{figure}

A detailed solution of the $e(t)$ equation is not needed to appreciate that circular rings subject to the forcing term in Eqn.~\ref{eqn:ring-forcing} are unstable to small-amplitude perturbations. When $0 < e \ll 1$, the ring particles near the far turnaround points (overlapping ellipses in \fref{fig:diagram}) experience a weak retrograde impulse, corresponding to the minidisk-minidisk impact. Backwards forcing near apocenter drives an angular momentum deficit, i.e. it increases the particle eccentricity, and that effect is not compensated around the minidisk pericenter because of the factor $\delta(\nu - \pi)$. The larger $e$ leads to a stronger retrograde impulse via Eqn. \ref{eqn:ring-forcing}, completing a feedback loop in which $e(t)$ grows exponentially. In \sref{sec:convergence} we determined the growth rate to be $\simeq 0.07 f_{\rm bin}$; this rate empirically fixes the constant in Eqn.~\ref{eqn:ring-forcing}.

\begin{figure} 
\centering
\includegraphics[width=0.47\textwidth]{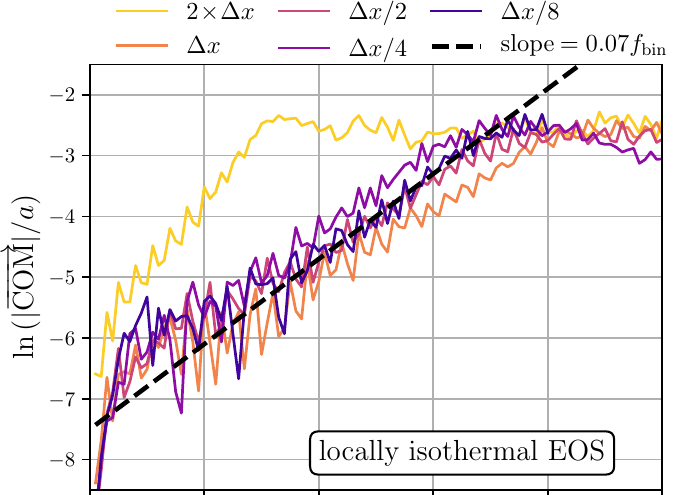}\\
\includegraphics[width=0.47\textwidth]{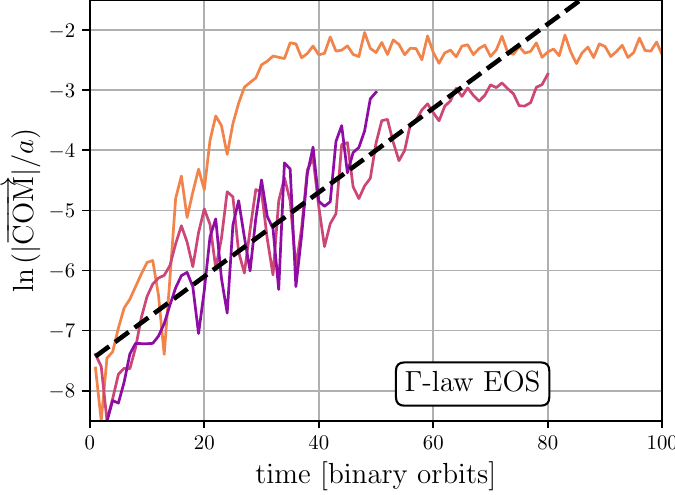}
\caption{Resolution study of the exponential growth phase of minidisk COM amplitudes for decretion, with the locally isothermal (top panel) and radiatively cooled $\Gamma$-law (bottom panel) equations of state. The fiducial resolution is $\Delta x=0.0025\,a$. The higher resolution $\Gamma$-law runs were too expensive to evolve to saturation.}
\label{fig:dec}
\end{figure}

A stochastic forcing term could be added as a model of gas falling in from the CBD and impacting the minidsks. The result should be the appearance of a non-zero but random-walking minidisk eccentricity, like what we observed in the simulations from \sref{sec:interaction} and \sref{sec:dephole}, where the minidisk-minidisk interaction was suppressed by use of a large absorber, or a ``hole'', respectively.

The mechanism proposed here for the eccentric minidisk instability does not deal with the hydrodynamical energy budget, nor the complicating effects of feeding from a CBD, and therefore cannot account for the instability's apparent sensitivity to the thermodynamical treatment when a CBD is present. Our numerical results are consistent with two possible interpretations. The first, is that the regularity of minidisk-minidisk impacts is compromised by the appearance of spiral arms, and that spiral arm formation is directly sensitive to the equation of state. The second, is that the equation of state directly affects the CBD morphology, which in turn sets the cadence and regularity of minidisk feeding, to which the eccentricity evolution is sensitive. In the second scenario, the absence of spiral arms (\fref{fig:MD_tiles}) could be a red herring, or it could be a consequence of the disks already being eccentric. This issue needs to be investigated further.

\subsection{Comparison to other eccentricity mechanisms} \label{sec:other}
The physical picture just proposed is adapted from one that was described in \citet[henceforth L94]{lubow1994} to explain the growth of eccentric disks around the white dwarf accretors of SU Ursae Majoris (SU UMa) binary systems. Those systems are seen to exhibit so-called ``superhump'' oscillations during periods of enhanced mass transfer from the donor star. The oscillations are widely interpreted as signifying an eccentric disk around the white dwarf, which precesses and causes the observed superhump mode to occur at the beat frequency with the binary orbital period. Eccentricity is known to be excited by the 3:1 Lindblad resonance \citep[e.g.][]{whitehurst+1991, lubow1991a, franchini+2019} operating in the outer edge of the disk, however L94 was exploring an alternative in which the eccentricity was driven instead by the gas stream from the donor star impacting the disk around the white dwarf.

The ballistic particle-ring approximation with external forcing was used in L94 to analyze the eccentricity injection by the impacting gas stream, however with a different forcing term from the one in Eqn.~\ref{eqn:ring-forcing}. In L94, the forcing strength was set proportional to the rate of mass flow from the donor star, which would not be in resonance with any waves excited in the disk. L94 showed that the ring eccentricity is excited during periods of increasing mass flow, but then is dissipated after the mass flow rate stabilizes to a new level. It was concluded in L94 that stream impacts were not a viable scenario for eccentricity injection in SU UMa systems, and that the 3:1 resonance was the more likely culprit.

\begin{figure}
\includegraphics{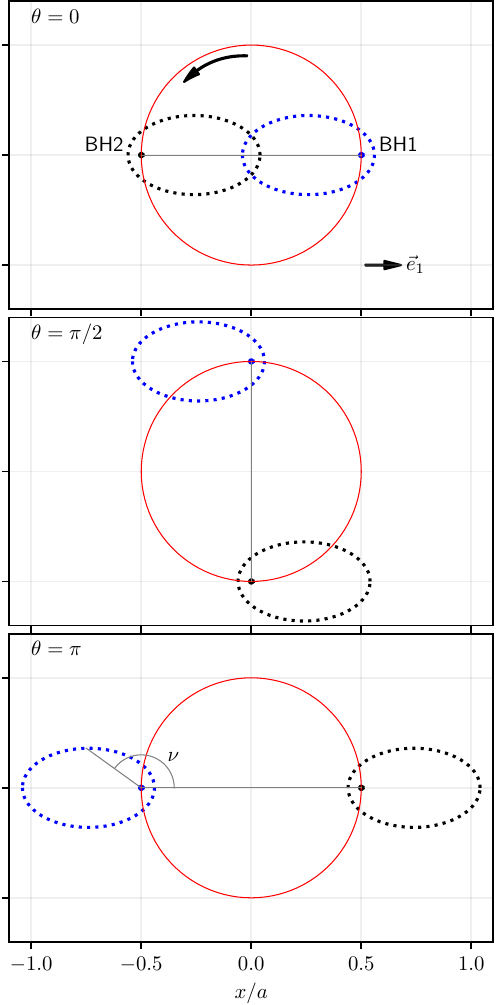}
\caption{The geometry of the minidisk-minidisk impacts. Dotted lines depict the outer edges of the minidisks, with the first black hole (BH1) in blue and the second (BH2) in black. The coordinate system is chosen so the minidisk apsides are horizontal, and the binary has orbital phase $\theta=0$ when BH1's minidisk's eccentricity vector points to the right (top panel). The red circle depicts the binary orbit. The true anomaly $\nu$ of a ring particle orbiting BH1 is shown in the bottom panel. The minidisks do not precess here, i.e.~their orientations are fixed in the inertial frame; the middle and bottom panels show the configuration at orbital phases $\theta=\pi/2$ and $\theta=\pi$ respectively. The minidisks collide and eccentricity is injected to the particle rings around the orbital phase $\theta=0$ (top panel).} \label{fig:diagram}
\end{figure}

It is relevant to note that we considered the 3:1 Lindblad resonance as a possible mechanism for the eccentric minidisk instability. However, that has been shown to succeed only when the binary mass ratio is $q \lesssim 1/3$ \citep[see e.g.][]{whitehurst+1991,murray+2000}, whereas we see the eccentric minidisk instability operating when $q=1$. Besides, the Lindblad resonance is a tidal interaction, and our results from \sref{sec:interaction} and \sref{sec:dephole} indicate the eccentric minidisk instability is being driven by resonant mass exchange.

In \sref{sec:interaction} we established that the resonant interaction could be destroyed by replacing one minidisk with a large absorber, but we also saw that some eccentricity was nonetheless developing in the extant minidisk, albeit without the coherent directionality. We interpreted this as arising from stochastic eccentricity injection by gas infall from the CBD, however we have considered the possibility that minidisks might also be susceptible to some kind of secular instability. 
For example, Kozai-Lidov oscillations can cause an initially circular minidisk to develop eccentricity \citep[][]{martin+2014, franchini+2019b}, but this mechanism does not apply in our case because it requires a large disk inclination with respect to the binary orbital plane.
In another example, isolated $\alpha$-disks were found by \cite{lyubarskij+1994} to be unstable to eccentricity growth by a viscous overstability. Later work by \cite{ogilvie2001} pointed out that viscous overstability could be an unphysical aspect of the $\alpha$-disk model, because it is suppressed when accounting for the finite relaxation time of magnetohydrodynamic turbulence expected in accretion disks. Possible scenarios where viscous overstability may be physical were further elucidated in \cite{latter+2006}. \cite{ogilvie2001} also showed that bulk viscosity suppresses viscous overstability, and this fact was used by \cite{kley+2008} to test whether viscous overstability was important in the development of eccentric disks.

A simple way to assess the importance of viscous overstability is to perform runs with non-zero kinematic bulk viscosity $\lambda$. Indeed, it was argued in \cite{kley+2008} that, if viscous overstability were important, then using $\lambda = 2 \nu$ would suppress disk eccentricity. We checked this case (see the panel labeled ``$\lambda \!=\! 2\nu\! =\! 10^{-4}$'' in \fref{fig:MD_tiles}), but we found no significant suppression of eccentricity (minidisk COM amplitude is still $\simeq 0.09\,a$). \cite{ogilvie2001} also derived a quenching condition for viscous overstability when $\alpha=0.1$, namely that the bulk $\alpha$-viscosity parameter is $>0.35$. Thus, we also checked a case with $\lambda/\nu > 0.35$ (see the panel labeled ``$\lambda = 4\nu = 10^{-4}$'' in \fref{fig:MD_tiles}), and we again found no suppression of minidisk eccentricity (minidisk COM amplitude is still $\simeq 0.09\,a$). In both bulk viscosity tests, we reduced $\nu$ to below $10^{-4}\sqrt{GMa}$ to ensure the tests were not affected by the viscous suppression demonstrated in the bottom panel of \fref{fig:COMvsnuMach}. We conclude that viscous overstability is not a likely explanation for the appearance of eccentric minidisks.

\subsection{Why gravitational softening produces less eccentric minidisks} \label{sec:softening}
We found (top row of \fref{fig:MD_tiles}) that minidisk eccentricity is suppressed by gravitational softening. This can be understood in terms of ballistic particle trajectories in softened gravitational potentials. Consider the effective potential for a gas parcel of specific angular momentum $\ell$ orbiting in the softened potential of an object with mass $M$,
\begin{equation*}
u_{\rm eff}(r) = \frac{\ell^2}{2 r^2} - \frac{G M}{\sqrt{r^2 + r_{\rm soft}^2}} \, .
\end{equation*}
The turning points in this potential are fixed by the specific orbital energy ${\cal E}$ of the gas parcel. Orbital eccentricity is not defined in the usual sense when $r_{\rm soft} > 0$, however, if $\ell$ and ${\cal E}$ are both fixed, then the radial distance between the turning points can be easily seen to decrease with increasing $r_{\rm soft}$. The result is that a given forcing amplitude (i.e.~a fixed value of the constant in Eqn.~\ref{eqn:ring-forcing}) results in a smaller geometrical distortion of the disk when $r_{\rm soft}$ is larger. This effect could be accounted for by replacing $e(t)$ in Eqn.~\ref{eqn:ring-forcing} with a different function that reflects the degree to which a ring with parameters $\ell$ and ${\cal E}$ is non-circular. Doing so would predict a slower growth rate and could account for the observed reduction of minidisk eccentricity with larger $r_{\rm soft}$.

\subsection{Softening-driven apsidal precession} \label{sec:precess}
In the vertically integrated thin disk setting, gravity is softened at second order in $h/r$, where $h$ is the vertical disk height measured from the midplane. This can be seen by introducing an ansatz for the vertical density profile, say $\rho = \rho_0 (1 - (z/h)^2)$ for $|z|<h$, $\rho=0$ otherwise, where $\rho_0$ has no dependence on $z$. The amplitude of the horizontal component of the gravitational force density is $\rho (GM/R^3) r$, where $R^2 = z^2 + r^2$ and $r$ is the cylindrical radial coordinate. Taking advantage of the fact that $z/r\ll 1$, we can write
\begin{equation}
    \rho \frac{GM}{R^3} r = \rho \frac{GM}{r^2} \left[ 1 - \frac{3}{2} \left(\frac{z}{r}\right)^2 \right] + \mathcal{O}\left(\frac{z}{r}\right)^4\!\!.
\end{equation}
Integrating over $z\in[-h,h]$ and defining $\Sigma \equiv \int_{-h}^h\! \rho\, dz$ yields
\begin{equation}
    \int_{-h}^h\!\!\!\! dz\, \rho \frac{GM}{R^3}r \simeq \Sigma \frac{GM}{r^2} \left[1 - \frac{3}{10} \left(\frac{h}{r}\right)^2 \right]\!.
\end{equation}
The factor in square brackets is $\leq 1$, and therefore weakens (``softens'') the gravitational force. Gravitational softening can thus be understood as modeling the finite thickness of the disk. If the factor in square brackets also decreases for decreasing $r$ (a condition which depends upon $h(r)$), it will soften more at smaller $r$, similar to the Plummer potential we use to model gravity in our simulations.

In practice, a commonly used model for softened gravity in thin disks derives from the Plummer potential \citep[][]{plummer1911},
\begin{equation}
    \Phi = -\frac{GM}{\sqrt{r^2 + r_{\rm soft}^2}}, \label{eq:plummer}
\end{equation}
where $r_{\rm soft}$ is the softening length. Based on comparisons to three-dimensional disks, the softening length in the Plummer potential ought to be on the order of the disk scale height \citep[][]{Mueller+2012}.\footnote{Note that any dependence of the softening length on the fluid variables or horizontal coordinates is usually ignored when taking the gradient of the Plummer potential.}

In this section, based on the Plummer model of gravitational softening, we describe conditions under which one might expect softening-driven retrograde apsidal precession of planar eccentric disks. To understand the role of softening, we consider a single gravitating mass, and neglect the disk self-gravity and hydrodynamic effects such as pressure gradients and effective viscosity \citep[see a discussion of these other effects in][]{Goodchild+2006}. We therefore operate in the ballistic approximation around a single gravitating mass, whereby fluid elements are treated as test masses moving freely under gravity. We take the softening length to be linear in $h$ with a constant of proportionality that is of order unity \citep[][]{Mueller+2012}.

Consider first the case of a razor thin disk, such that $r_{\rm soft} \propto h = 0$. In this case, the gravitational force is Newtonian, thus we expect eccentric orbits to be closed ellipses (i.e.~zero precession). 
The same expectation holds whenever the disk has a constant aspect ratio, $h \!\propto\! r$, because then $r_{\rm soft}\! \propto\! r$ and the Plummer potential becomes proportional to the Newtonian potential. In this case, eccentric orbits are still closed ellipses, but it is as though the central gravitating object has a suppressed mass. This condition is representative of gas-dominated $\alpha$-disks, as they have relatively constant aspect ratios \citep[since $h \propto r^{21/20}$ or $h \propto r^{9/8}$ when electron or free-free scattering opacity dominate, respectively; see e.g.][]{Haiman:2009}.

On the other hand, radiation-dominated disks have constant disk scale height \citep[see e.g.][]{Haiman:2009}, i.e.~$h/r \sim 1/r$, so that the Plummer force is approximately Newtonian for $r\!\gg\! h$ but weaker than Newtonian for $r\! \simeq\! h$. 
In this case, the deficit of (vertically integrated) gravity near the central object causes eccentric orbits to precess in the retrograde direction. This can be understood intuitively as follows. At large distances, the eccentric orbit is approximately Newtonian, i.e.~an ellipse. But close to the central object, gravity becomes increasingly weaker than Newtonian, and unable to close the particle trajectory to an ellipse. This causes its next apocenter to be rotated in the direction opposite of the orbital motion. We verified this picture numerically by evolving test particle trajectories in a Plummer potential.

\subsection{Eccentric precessing minidisks in 2D versus 3D} \label{sec:pars}
As guidance for three-dimensional studies, in this section we point to regions of parameter space where the effects of minidisk eccentricity and retrograde precession may reveal themselves. 

Since minidisk eccentricity is triggered by mass-trading activity between minidisks, it is important that such activity not be disrupted by, e.g.~artificial obstructions between them. Thus, the entire region between the binary should be resolved. Since minidisk eccentricity is suppressed by viscosity, three-dimensional studies seeking to reveal eccentric minidisks should have weaker effective viscosity. The value of $\alpha=0.01$ achieved in~\cite{oyang+2021}, for example, should be amply low, since we find eccentric minidisks with viscosity as high as $\alpha=0.1$. Since minidisk eccentricity is suppressed by gravitational softening, and softening represents the finite thickness of disks, a three-dimensional investigation should strive to make the disk thinner. Simulating thinner disks in three dimensions increases computational cost due to the higher resolution required in the $z$-direction. Although this does not increase the number of cells required in the vertical direction, the cells are smaller, which tightens the time step constraint. If one strives to simulate the $r_{\rm soft}=0.01\,a$ case (which yields obvious minidisk eccentricity in 2 dimensions, e.g.~$e\sim 0.5$), and assuming the softening length is $\simeq 0.5\, h$ where $h$ is the disk half-thickness, then one requires that $h$ does not exceed $\simeq 0.02\,a$ within the minidisks. Note that the insensitivity of minidisk eccentricity to Mach number in our study is not expected to be reproduced in three-dimensional studies, because the effective softening length is intimately tied to Mach number via $\mathcal{M} \sim (h/r)^{-1}$; whereas in our study, the softening length is instead an independent ad hoc parameter, artificially decoupled from Mach number.

On the other hand, testing softening-driven retrograde precession in three-dimensional studies requires disks that are \emph{sufficiently} thin (such that minidisk eccentricity is appreciable), but still sufficiently thick that the effect of the implied gravitational softening on precession dominates over other hydrodynamical and tidal effects (see \sref{sec:masstrade}). Our results suggest that the $r_{\rm soft}=0.01\,a$ case should be sufficient (i.e.~$h\simeq 0.02\,a$). However, the functional form of the Plummer potential also suggests that disks with nearly constant aspect ratios will not undergo retrograde precession (see \sref{sec:precess}); instead, three-dimensional studies seeking to reveal retrograde minidisk precession should focus on flatter disk profiles (such as constant disk heights expected in radiation-dominated disks). Note that retrograde precession should not require a binary, so a targeted simulation of a disk with flat height profile and eccentricity initialized to e.g.~$e\simeq0.5$ around a single gravitating object ought to be a sufficient test of the effect. Cylindrical polar coordinates, rather than spherical, would be efficient for simulations of flat disks.

In their relativistic simulations, \cite{avara+2023} reported time-varying tilts of the minidisks out of the equatorial plane, with tilt angles comparable to the aspect ratio of the disk. If severe enough, such tilts could cause eccentric minidisks to miss each other at the phase $\theta=0$ shown in \fref{fig:diagram}, thereby inhibiting eccentricity growth. Thus, it is conceivable that three-dimensional effects not captured in the vertically integrated approach could prevent minidisk eccentricity growth in some scenarios.

There are subtleties about our two-dimensional models which may be important to understand when comparing with three-dimensional simulations. Firstly, the Plummer potential is an ad hoc model of the gravitational softening that occurs when integrating out the vertical degree of freedom in thin disks. In particular, it is not derived in a controlled perturbative procedure in powers of the local disk aspect ratio. To do so would require greater knowledge of the local vertical density profile. Thus, a lack of softening-driven retrograde precession in constant aspect ratio disks is only predicted to the extent that the Plummer potential is a reasonable model of softened gravity in that regime \citep[see][for some validations of the Plummer potential against three-dimensional calculations]{Mueller+2012}. However, we expect that retrograde precession in flat disks (i.e.~$h = \:$constant) should be a generic consequence of gravitational softening, independent of the applicability of the Plummer model.

Secondly, when performing a vertical integration of the hydrodynamic equations of a thin disk, the gravitational force softens beginning at second order in the disk aspect ratio. Instead, if a polar integration is performed, the magnitude of the gravitational force per unit area can be calculated at fully nonlinear order as $GM\Sigma_{\rm polar}/R^2$ since the coordinates conform with the spherical symmetry of the point mass gravitational potential. Here, we defined $\Sigma_{\rm polar} \equiv \int_{\pi/2-\theta_h}^{\pi/2+\theta_h} \rho R d\theta$ to be the exact surface density, where $\theta_h$ defines the disk's local polar extent measured from the midplane. In other words, with polar integration, gravity does not appear to have a softened functional form at fully nonlinear order. Thus, it is more cautious and nuanced to say that retrograde precession is possibly a finite-thickness effect, which manifests via softened gravity under vertical integration, but may have alternative physical interpretations in other two-dimensional reductions. Ambiguity in the physical interpretation of thin disks was recognized in~\cite{Abramowicz+1997}, e.g.~the use of spherical versus cylindrical coordinates trades between a vertical gravitational force and a vertical centrifugal force. On this note, it is worth pointing out that pressure and finite disk thickness (which are not mutually exclusive) are known to influence the apsidal precession rates of disks \citep[see e.g.][]{kato1983, Goodchild+2006}, with pressure effects in particular giving rise to retrograde forcing (as long as radial derivatives of pressure are not too positive) which becomes stronger for thicker disks \citep[][]{Goodchild+2006}. This increased retrograde driving with thicker disks \citep[also reported in][]{kley+2008} is at least directionally consistent with the softening-driven precession we describe in this work (i.e.~thicker disks imply larger softening, which implies stronger retrograde driving).

Lastly, two-dimensional models of disks with explicit viscosity are, strictly speaking, turbulence closure models. Hence, the evolved variables must be understood as suitable averages \citep[e.g.~Favre-averaged quantities, see][]{favre1969}. A strict comparison with three-dimensional simulations would therefore necessitate computing such averaged quantities. If eccentric minidisks and softening-driven retrograde precession are manifested in three dimensions in an average sense, but not in any instantaneous sense, or if these effects are complete artifacts of the two-dimensional models, this would be a peculiar aspect of turbulence models of thin disks that theorists ought to be aware of.

\subsection{Observable consequences} \label{sec:obs}
\begin{figure} 
\centering
\includegraphics[width=0.47\textwidth]{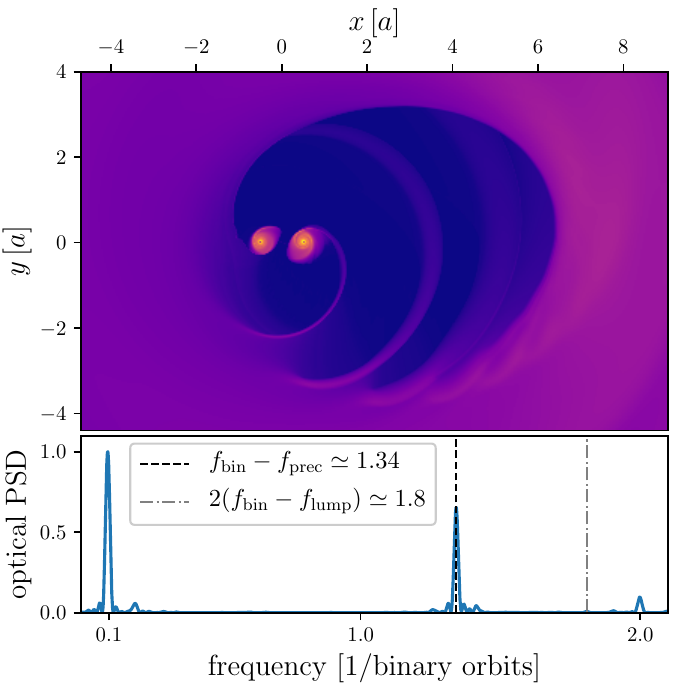}
\caption{Top: Images of the scaled surface density $\Sigma^{1/3}$, showing the extent of the eccentric cavity forming in radiatively cooled $\Gamma$-law runs. Bottom: Normalized PSD of the optical light curve for the same system as in \fref{fig:MDCOM}. The peak around $0.1 f_{\rm bin}$ is the lump frequency and the peak at $1.34 f_{\rm bin}$ is the ``eccentric minidisk beat frequency'' $f_{\rm bin} - f_{\rm prec}$. The large cavity leads to suppression of the ``binary-lump beat frequency'' $2(f_{\rm bin} - f_{\rm lump})$ introduced in \sref{sec:background}.}
\label{fig:beats}
\end{figure}

In our runs varying the softening length (labeled S0 -- S5 in Table~\ref{tab:simsuite}, and see the first row of \fref{fig:MD_tiles}), $\sim 2-13\%$ of the minidisk mass is exchanged between minidisks per trading event (see the red curve in \fref{fig:COMvsrsoft}). Averaged over time, this corresponds to a mass exchange rate of $\sim 0.6-2.1$ times the total mass accretion rate (see \fref{fig:mdot_vs_mtrade}). Mass trading events can therefore be significant hydrodynamic events that cause observable EM flares. 
In our previous work W22, we reported simulated light curves \footnote{Note that the simulations presented here, as well as in W22 and in many similar previous studies, predict only the (mainly UV) thermal emission from the optically thick plasma, heated by viscosity and also shocks. At higher photon energies, additional non-thermal radiation is expected to be produced, similar to the hot coronal X-ray emission observed in AGN powered by solitary SMBHs \citep[see e.g.][]{sesana+2012}.} from accreting equal-mass circular binaries which exhibit periodic flaring at near-orbital frequency. In this work, we explain this periodicity as corresponding to mass trading events between minidisks. Flares are caused by shock-heating during minidisk-minidisk collisions, and we have checked that the flares are coincident in time with these collisions; we leave a more detailed study of each collision to future work. The frequency of mass trading is a beat frequency $f_{\rm bin} - f_{\rm prec}$ between the binary orbit $f_{\rm bin}$ and the signed minidisk precession $f_{\rm prec}$ (negative values meaning retrograde precession), indicated with the dotted vertical line in the bottom panel of \fref{fig:beats}. The system's optical periodogram (obtained in the same way as W22) exhibits a peak at this beat frequency. 

The well-established $m=1$ overdensity in the circumbinary disk (called the ``lump'') has a pattern speed which imprints on the optical emission at a frequency of $f_{\rm lump} = 0.1$ per binary orbit (see the leftmost peak in \fref{fig:beats}). We note that $f_{\rm bin} - f_{\rm prec}$ is a distinct physical phenomenon from the beat frequency between the binary and the lump, $2(f_{\rm bin} - f_{\rm lump})$ (shown as the dash-dotted vertical line in \fref{fig:beats}), which forms when all sides of the cavity wall are sufficiently close to the binary that the binary can strip material from it at almost all lump phases \citep[e.g.~cavities as reported in][which are smaller, less offset, and less eccentric than the run we show in \fref{fig:beats}]{noble+2012, bowen_lump, Gutierrez+2021, combi+2021}. The top panel of \fref{fig:beats} shows a snapshot of $\Sigma$ (raised to the $1/3$rd power to improve contrast), showing that the cavity is far too large and offset for the binary-lump beat frequency to form.

Also note that a similar minidisk mass-exchange phenomenon was observed in relativistic simulations \citep[][]{bowen+2017,avara+2023}. That phenomenon was reported as being an effect enhanced to a significant level by relativistic gravity, characterized by a sloshing flow behavior resulting mostly in alternating mass transfer between minidisks, and associated with large morphological transitions of the minidisks \citep[referred to as ``disk-dominated'' and ``stream-dominated'' states in][]{avara+2023}. This contrasts with our finding in many ways, since we find a strong phenomenon occurring even at the Newtonian level, characterized by eccentric, precessing minidisks, and sychronized mass trading events. Assuming the phenomenon we report in this work is indeed distinct from that reported in \cite{bowen+2017} and \cite{avara+2023}, there is potential to confuse their observational signatures. The cadence of the sloshing effect between minidisks reported in the more recent work by \cite{avara+2023} is roughly $1.4\times$ per orbit, quite similar to our $f_{\rm bin} - f_{\rm prec}$ when the softening length is $\sim 0.04\,a$ (see \fref{fig:rsoftlim}). 

The rate of mass exchanged in the sloshing mechanism \citep[as reported in][]{avara+2023}, is at the level of $0.1 \times \dot{M}_{\rm BHs}$, whereas the eccentric minidisk instability can lead to a much larger mass exchange rate of $\sim\,0.6-2.1 \times \dot{M}_{\rm BHs}$ (e.g.~see our run S1 in \fref{fig:mdot_vs_mtrade}).

If eccentric minidisks do form in accreting SMBHBs, we showed in W22 that they could produce a detectable QPO at or near the binary orbital period, likely in the UV or optical bands. This knowledge could aid in the identification of EM counterparts to future individual-source detections by the pulsar timing arrays, or assist in targeted searches by placing a prior on the binary orbital period given the QPO periodicity \cite[e.g.][]{Arzoumanian2020}.

\section{Conclusions \& Outlook} \label{sec:conclude}
In this work, we showed that accreting, circular, equal mass binaries are prone to an instability which grows a significant eccentricity in the minidisks around the binary components. The mechanism originates in mass trading between the minidisks, which tends to synchronize and become periodic, driving up eccentricity, and causing the eccentric minidisks to maintain opposing orientations. This process is especially strong in the limit of small gravitational softening. Gas impacts from a circumbinary disk are neither necessary nor sufficient to explain this effect.

We investigated the dependence of minidisk eccentricity on model details. We found that many model choices, such as the use of artificial target temperature profiles (e.g.~the use of $\beta$-cooling or locally isothermal equations of state), large gravitational softening (e.g.~$r_{\rm soft} = 0.05\,a$), large viscosity (e.g.~$\nu = 10^{-3} \sqrt{GMa}$), and a grid obstruction between the minidisks, all suppress minidisk eccentricity. This may partly explain why significant minidisk eccentricity in circular equal-mass binaries has not been previously reported in the literature. We found that minidisk eccentricity is robust to large bulk-to-shear viscosity ratios, which suggests this phenomenon is robust to a finite relaxation time of magnetohydrodynamic turbulence \citep[][]{ogilvie2001}.

We also showed that eccentric minidisks tend to precess steadily in the retrograde direction when gravity is softened. In the limit of zero softening, the precession can in general be prograde, zero, or retrograde, depending on the balance of driving from hydrodynamic and tidal forces. The minidisks trade mass at a beat frequency, $f_{\rm bin} - f_{\rm prec}$, between the binary orbital frequency $f_{\rm bin}$ and the minidisk precession frequency $f_{\rm prec}$; note the minidisk precession frequency is negative when the precession is retrograde. This ``eccentric minidisk beat frequency'' imprints on light curves from thermal disk emission, as we reported in~\cite{WS2022}; in this work we clarified that the physical origin of such periodicity is the minidisk mass trade. Although the frequency can be similar, this effect is distinct from the ``binary-lump beat frequency'' $2(f_{\rm bin} - f_{\rm lump})$ formed between the lump and the binary, which occurs when the cavity wall is sufficiently close to the binary that the binary can draw material from the lump at most lump phases.

In a careful interpretation of the two-dimensional thin disk setting, we argued that softening-driven retrograde precession is a finite-thickness effect, even though a precise physical interpretation is not obvious. We believe that future three-dimensional simulations could observe the eccentric minidisk instability, but acknowledge that high resolution may be required due to the possible need for a rather thin disk, with $h/r \lesssim 0.02$. Three-dimensional simulations could also help clarify the physical meaning of gravitational softening commonly used in vertically integrated hydrodynamical settings. Seeing as we restricted this work to circular, equal-mass binaries, future work should also determine the range of binary parameters where this eccentric minidisk instability operates.

\begin{acknowledgements}
All simulations were performed on Clemson University's Palmetto cluster, and we gratefully acknowledge the Palmetto HPC support team. We acknowledge support from National Science Foundation grants AST-2006176 (to ZH) and AST-1715661 (to ZH and AM), NASA ATP grant 80NSSC22K0822 (to ZH and AM), and use of the software NumPy \citep{numpy}, Matplotlib \citep{Hunter:2007}, SciPy \citep{scipy}, CuPy \citep{cupy}. We thank the KITP at UC Santa Barbara for their hospitality during the Binary22 workshop, where some of the early work for this project was performed. We also acknowledge Julian Krolik, Mark Avara, Alessia Franchini, Matthew Bate, and Steve Lubow for valuable discussions at that workshop and since. KITP is supported in part by the National Science Foundation under Grant No. NSF PHY-1748958.
\end{acknowledgements}

\appendix

\section{Simulation Suite} \label{app:simsuite}

\begin{deluxetable}{lllllllll} 
\tablecaption{Summary of our simulation suite}

\tablehead{\colhead{$\!$Label$\!$} & \colhead{$\! r_{\rm soft}\,[a]\!$} & \colhead{$\! s \,[\Omega_{\rm bin}]\!$} & \colhead{$\mathcal{M}$} & \multicolumn{1}{l}{Grid refinement} & \multicolumn{1}{l}{Viscosity} & \multicolumn{1}{l}{Cooling} & \colhead{$\! r_{\rm sink} [a]\!$} & \multicolumn{1}{l}{Special condition} }

\startdata
S0  & 0 & 32       & 11    & \multicolumn{1}{p{3.5cm}}{$0.012\,a$ to $1000\, T_{\rm bin}$, then\newline$0.006\,a$ to $1100\, T_{\rm bin}$, then\newline$0.003\,a$ to the final time} & $\alpha=0.1$   & \multicolumn{1}{l}{Radiative}  & 0.02 & \multicolumn{1}{p{2.2cm}}{Soften inside{\newline}sink only}         \\ 
S1  & \multicolumn{1}{l}{0.01} & \multicolumn{1}{l}{16}   & \multicolumn{1}{l}{11}    & \multicolumn{1}{l}{As above}           & \multicolumn{1}{l}{$\alpha=0.1$}   & \multicolumn{1}{l}{Radiative}  & \multicolumn{1}{l}{0.02} & \multicolumn{1}{l}{\ldots}         \\ 
S2  & \multicolumn{1}{l}{0.02} & \multicolumn{1}{l}{8}    & \multicolumn{1}{l}{11}    & \multicolumn{1}{p{3.5cm}}{$0.012\,a$ to $1000\, T_{\rm bin}$, then\newline$0.006\,a$ to the final time}           & \multicolumn{1}{l}{$\alpha=0.1$}   & \multicolumn{1}{l}{Radiative}  & \multicolumn{1}{l}{0.02} & \multicolumn{1}{l}{\ldots}         \\ 
S3  & \multicolumn{1}{l}{0.03} & \multicolumn{1}{l}{8}    & \multicolumn{1}{l}{11}    & \multicolumn{1}{l}{As above}           & \multicolumn{1}{l}{$\alpha=0.1$}   & \multicolumn{1}{l}{Radiative}  & \multicolumn{1}{l}{0.02} & \multicolumn{1}{l}{\ldots}         \\ 
S4  & \multicolumn{1}{l}{0.04} & \multicolumn{1}{l}{8}    & \multicolumn{1}{l}{11}    & \multicolumn{1}{l}{As above}           & \multicolumn{1}{l}{$\alpha=0.1$}   & \multicolumn{1}{l}{Radiative}  & \multicolumn{1}{l}{0.02} & \multicolumn{1}{l}{\ldots}         \\ 
S5  & \multicolumn{1}{l}{0.05} & \multicolumn{1}{l}{8}    & \multicolumn{1}{l}{11}    & \multicolumn{1}{l}{As above}           & \multicolumn{1}{l}{$\alpha=0.1$}   & \multicolumn{1}{l}{Radiative}  & \multicolumn{1}{l}{0.02} & \multicolumn{1}{l}{\ldots}         \\ 
M7  & \multicolumn{1}{l}{0.02} & \multicolumn{1}{l}{8}    & \multicolumn{1}{l}{7}    & \multicolumn{1}{p{3.5cm}}{$0.024\,a$ to $3400\, T_{\rm bin}$, then\newline$0.012\,a$ to $3500\,T_{\rm bin}$, then\newline$0.006\,a$ to the final time}           & \multicolumn{1}{l}{$\nu=10^{-4}$}   & \multicolumn{1}{l}{Radiative}  & \multicolumn{1}{l}{0.02} & \multicolumn{1}{l}{\ldots}\\ 
M11  & \multicolumn{1}{l}{0.02} & \multicolumn{1}{l}{8}       & \multicolumn{1}{l}{11}    & \multicolumn{1}{l}{As above}           & \multicolumn{1}{l}{$\nu=10^{-4}$}   & \multicolumn{1}{l}{Radiative}  & \multicolumn{1}{l}{0.02} & \multicolumn{1}{l}{\ldots}\\ 
M15  & \multicolumn{1}{l}{0.02} & \multicolumn{1}{l}{8}       & \multicolumn{1}{l}{15}    & \multicolumn{1}{l}{As above}           & \multicolumn{1}{l}{$\nu=10^{-4}$}   & \multicolumn{1}{l}{Radiative}  & \multicolumn{1}{l}{0.02} & \multicolumn{1}{l}{\ldots}\\ 
M20  & \multicolumn{1}{l}{0.02} & \multicolumn{1}{l}{8}       & \multicolumn{1}{l}{20}    & \multicolumn{1}{l}{As above}           & \multicolumn{1}{l}{$\nu=10^{-4}$}   & \multicolumn{1}{l}{Radiative}  & \multicolumn{1}{l}{0.02} & \multicolumn{1}{l}{\ldots}\\ 
M25  & \multicolumn{1}{l}{0.02} & \multicolumn{1}{l}{8}       & \multicolumn{1}{l}{25}    & \multicolumn{1}{l}{As above}           & \multicolumn{1}{l}{$\nu=10^{-4}$}   & \multicolumn{1}{l}{Radiative}  & \multicolumn{1}{l}{0.02} & \multicolumn{1}{l}{\ldots}\\ 
V1  & \multicolumn{1}{l}{0.02} & \multicolumn{1}{l}{8}       & \multicolumn{1}{l}{11}    & \multicolumn{1}{l}{As above}           & \multicolumn{1}{l}{$\nu=10^{-4}$}   & \multicolumn{1}{l}{Radiative}  & \multicolumn{1}{l}{0.02} & \multicolumn{1}{l}{\ldots}
\\ 
V4  & \multicolumn{1}{l}{0.02} & \multicolumn{1}{l}{8}       & \multicolumn{1}{l}{11}    & \multicolumn{1}{l}{As above}           & \multicolumn{1}{l}{$\nu=4\!\times\!10^{-4}$}   & \multicolumn{1}{l}{Radiative}  & \multicolumn{1}{l}{0.02} & \multicolumn{1}{l}{\ldots}
\\ 
V7  & \multicolumn{1}{l}{0.02} & \multicolumn{1}{l}{8}       & \multicolumn{1}{l}{11}    & \multicolumn{1}{l}{As above}           & \multicolumn{1}{l}{$\nu=7\!\times\!10^{-4}$}   & \multicolumn{1}{l}{Radiative}  & \multicolumn{1}{l}{0.02} & \multicolumn{1}{l}{\ldots}
\\ 
V10  & \multicolumn{1}{l}{0.02} & \multicolumn{1}{l}{8}       & \multicolumn{1}{l}{11}    & \multicolumn{1}{l}{As above}           & \multicolumn{1}{l}{$\nu=10^{-3}$}   & \multicolumn{1}{l}{Radiative}  & \multicolumn{1}{l}{0.02} & \multicolumn{1}{l}{\ldots} \\ 
BV2 & \multicolumn{1}{l}{0.02} & \multicolumn{1}{l}{8}       & \multicolumn{1}{l}{11}    & \multicolumn{1}{l}{As above}           & \multicolumn{1}{l}{$\lambda=2\nu=10^{-4}$}   & \multicolumn{1}{l}{Radiative}  & \multicolumn{1}{l}{0.02} & \multicolumn{1}{l}{\ldots} \\ 
BV4 & \multicolumn{1}{l}{0.02} & \multicolumn{1}{l}{8}       & \multicolumn{1}{l}{11}    & \multicolumn{1}{l}{As above}           & \multicolumn{1}{l}{$\lambda=4\nu=10^{-4}$}   & \multicolumn{1}{l}{Radiative}  & \multicolumn{1}{l}{0.02} & \multicolumn{1}{l}{\ldots} \\ 
$\beta$1  & \multicolumn{1}{l}{0.02} & \multicolumn{1}{l}{8}       & \multicolumn{1}{l}{10}    & \multicolumn{1}{l}{As above}           & \multicolumn{1}{l}{$\nu=10^{-4}$}   & \multicolumn{1}{p{1.7cm}}{$\beta=10$}  & \multicolumn{1}{l}{0.02} & \multicolumn{1}{l}{\ldots}
\\ 
$\beta$0  & \multicolumn{1}{l}{0.02} & \multicolumn{1}{l}{8}       & \multicolumn{1}{l}{10}    & \multicolumn{1}{l}{As above}           & \multicolumn{1}{l}{$\nu=10^{-4}$}   & \multicolumn{1}{p{1.7cm}}{$\beta=1$}  & \multicolumn{1}{l}{0.02} & \multicolumn{1}{l}{\ldots} \\ 
$\beta$m3  & \multicolumn{1}{l}{0.02} & \multicolumn{1}{l}{8}       & \multicolumn{1}{l}{10}    & \multicolumn{1}{l}{As above}           & \multicolumn{1}{l}{$\nu=10^{-4}$}   & \multicolumn{1}{p{1.7cm}}{$\beta=10^{-3}$}  & \multicolumn{1}{l}{0.02} & \multicolumn{1}{l}{\ldots} \\ 
ISO  & \multicolumn{1}{l}{0.02} & \multicolumn{1}{l}{8}       & \multicolumn{1}{l}{10}    & \multicolumn{1}{l}{As above}           & \multicolumn{1}{l}{$\nu=10^{-4}$}   & \multicolumn{1}{l}{Instantaneous}  & \multicolumn{1}{l}{0.02} & \multicolumn{1}{l}{Locally isothermal}
\\ 
HOLE  & \multicolumn{1}{l}{0.02} & \multicolumn{1}{l}{8}       & \multicolumn{1}{l}{11}    & \multicolumn{1}{p{3.5cm}}{$0.012\,a$ to $1000\, T_{\rm bin}$, then\newline$0.006\,a$ to the final time}           & \multicolumn{1}{l}{$\alpha=0.1$}   & \multicolumn{1}{l}{Radiative}  & \multicolumn{1}{l}{0.02} & \multicolumn{1}{p{2.2cm}}{3rd sink at\newline$r<0.05\,a$}
\\ 
H1  & \multicolumn{1}{l}{0.04} & \multicolumn{1}{l}{8}       & \multicolumn{1}{l}{11}    & \multicolumn{1}{p{3.5cm}}{$0.02\,a$ to $3000\, T_{\rm bin}$, then\newline$0.01\,a$ to $3100\,T_{\rm bin}$, then\newline$0.005\,a$ to $3200\,T_{\rm bin}$, then\newline$0.0025\,a$ to the final time}           & \multicolumn{1}{l}{$\alpha=0.1$}   & \multicolumn{1}{l}{Radiative}  & \multicolumn{1}{l}{0.01} & \multicolumn{1}{l}{\ldots}
\\ 
H2  & \multicolumn{1}{l}{0.01} & \multicolumn{1}{l}{8}       & \multicolumn{1}{l}{11}    & \multicolumn{1}{l}{As above}           & \multicolumn{1}{l}{$\alpha=0.1$}   & \multicolumn{1}{l}{Radiative}  & \multicolumn{1}{l}{0.01} & \multicolumn{1}{l}{Deplete CBD}
\\ 
H3  & \multicolumn{1}{l}{0.01} & \multicolumn{1}{l}{8}       & \multicolumn{1}{l}{11}    & \multicolumn{1}{l}{As above}           & \multicolumn{1}{l}{$\alpha=0.1$}   & \multicolumn{1}{l}{Radiative}  & \multicolumn{1}{l}{0.01} & \multicolumn{1}{l}{Refill MDs}
\\ 
H4  & \multicolumn{1}{l}{0.01} & \multicolumn{1}{l}{8}       & \multicolumn{1}{l}{11}    & \multicolumn{1}{l}{As above}           & \multicolumn{1}{l}{$\alpha=0.1$}   & \multicolumn{1}{l}{Radiative}  & \multicolumn{1}{l}{0.01} & \multicolumn{1}{l}{Absorb one MD}
\\ 
VH3  & \multicolumn{1}{l}{0.01} & \multicolumn{1}{l}{8}       & \multicolumn{1}{l}{11}    & \multicolumn{1}{l}{$0.00125\,a$ to the final time}           & \multicolumn{1}{l}{$\alpha=0.1$}   & \multicolumn{1}{l}{Radiative}  & \multicolumn{1}{l}{0.01} & \multicolumn{1}{l}{Zoom-in of H3}
\\ 
D0   & \multicolumn{1}{l}{0} & \multicolumn{1}{l}{\ldots}       & \multicolumn{1}{l}{\ldots}    & \multicolumn{1}{l}{$0.0025\,a$ to the final time$\!\!\!$}           & \multicolumn{1}{l}{$\alpha=0.001$}   & \multicolumn{1}{l}{Radiative}  & \multicolumn{1}{l}{0.02} & \multicolumn{1}{p{2.78cm}}{Decretion, no CBD,\newline soften inside\newline sink only}
\\ 
D1   & \multicolumn{1}{l}{0.01} & \multicolumn{1}{l}{\ldots}       & \multicolumn{1}{l}{\ldots}    & \multicolumn{1}{l}{$0.0025\,a$ to the final time$\!\!\!$}           & \multicolumn{1}{l}{$\alpha=0.1$}   & \multicolumn{1}{l}{Radiative}  & \multicolumn{1}{l}{0.02} & \multicolumn{1}{l}{Decretion, no CBD} \\
\enddata
\label{tab:simsuite}
  \vspace{6pt}
  {\bf Notes.} The quoted Mach number $\mathcal{M}$ is at $r\!=\!a$ in the initial conditions. The kinematic shear and bulk viscosities $\nu$ \& $\lambda$ have units of $\sqrt{GMa}$. Runs labeled S0-S5 vary the softening length; M7-M25 vary the Mach number; V1-V10 vary the kinematic shear viscosity; BV2, BV4 include bulk viscosity; $\beta$1, $\beta$0, and $\beta$m3 use target-temperature ``$\beta$'' cooling; ISO uses a locally isothermal EOS; HOLE has a sink at the origin; H1-H4 are high-resolution runs; VH3 is a very high-resolution version of H3; D0-D1 are decretion runs without a CBD; the mini decretion disks have Mach numbers of $\sim10$ for D0 and $\sim18$ for D1.
\end{deluxetable}

%
%
\renewcommand\bibname{References}

\bibliographystyle{aasjournal}
\bibliography{cbd}

\begin{thebibliography}{}
\expandafter\ifx\csname natexlab\endcsname\relax\def\natexlab#1{#1}\fi
\providecommand{\url}[1]{\href{#1}{#1}}
\providecommand{\dodoi}[1]{doi:~\href{http://doi.org/#1}{\nolinkurl{#1}}}
\providecommand{\doeprint}[1]{\href{http://ascl.net/#1}{\nolinkurl{http://ascl.net/#1}}}
\providecommand{\doarXiv}[1]{\href{https://arxiv.org/abs/#1}{\nolinkurl{https://arxiv.org/abs/#1}}}

\bibitem[{{Abramowicz} {et~al.}(1997){Abramowicz}, {Lanza}, \&
  {Percival}}]{Abramowicz+1997}
{Abramowicz}, M.~A., {Lanza}, A., \& {Percival}, M.~J. 1997, \apj, 479, 179,
  \dodoi{10.1086/303869}

\bibitem[{Afzal {et~al.}(2023)Afzal, Agazie, Anumarlapudi, Archibald,
  Arzoumanian, Baker, Bécsy, Blanco-Pillado, Blecha, Boddy, Brazier, Brook,
  Burke-Spolaor, Burnette, Case, Charisi, Chatterjee, Chatziioannou,
  Cheeseboro, Chen, Cohen, Cordes, Cornish, Crawford, Cromartie, Crowter,
  Cutler, DeCesar, DeGan, Demorest, Deng, Dolch, Drachler, von Eckardstein,
  Ferrara, Fiore, Fonseca, Freedman, Garver-Daniels, Gentile, Gersbach, Glaser,
  Good, Guertin, Gültekin, Hazboun, Hourihane, Islo, Jennings, Johnson, Jones,
  Kaiser, Kaplan, Kelley, Kerr, Key, Laal, Lam, Lamb, Lazio, Lee, Lewandowska,
  dos Santos, Littenberg, Liu, Lorimer, Luo, Lynch, Ma, Madison, McEwen, McKee,
  McLaughlin, McMann, Meyers, Meyers, Mingarelli, Mitridate, Nay, Natarajan,
  Ng, Nice, Ocker, Olum, Pennucci, Perera, Petrov, Pol, Radovan, Ransom, Ray,
  Romano, Sardesai, Schmiedekamp, Schmiedekamp, Schmitz, Schröder, Schult,
  Shapiro-Albert, Siemens, Simon, Siwek, Stairs, Stinebring, Stovall,
  Stratmann, Sun, Susobhanan, Swiggum, Taylor, Taylor, Trickle, Turner, Unal,
  Vallisneri, Verma, Vigeland, Wahl, Wang, Witt, Wright, Young, Zurek, \&
  Collaboration}]{Afzal2023}
Afzal, A., Agazie, G., Anumarlapudi, A., {et~al.} 2023, The Astrophysical
  Journal Letters, 951, L11, \dodoi{10.3847/2041-8213/acdc91}

\bibitem[{Agazie {et~al.}(2023)Agazie, Anumarlapudi, Archibald, Arzoumanian,
  Baker, Bécsy, Blecha, Brazier, Brook, Burke-Spolaor, Burnette, Case,
  Charisi, Chatterjee, Chatziioannou, Cheeseboro, Chen, Cohen, Cordes, Cornish,
  Crawford, Cromartie, Crowter, Cutler, DeCesar, DeGan, Demorest, Deng, Dolch,
  Drachler, Ellis, Ferrara, Fiore, Fonseca, Freedman, Garver-Daniels, Gentile,
  Gersbach, Glaser, Good, Gültekin, Hazboun, Hourihane, Islo, Jennings,
  Johnson, Jones, Kaiser, Kaplan, Kelley, Kerr, Key, Klein, Laal, Lam, Lamb,
  Lazio, Lewandowska, Littenberg, Liu, Lommen, Lorimer, Luo, Lynch, Ma,
  Madison, Mattson, McEwen, McKee, McLaughlin, McMann, Meyers, Meyers,
  Mingarelli, Mitridate, Natarajan, Ng, Nice, Ocker, Olum, Pennucci, Perera,
  Petrov, Pol, Radovan, Ransom, Ray, Romano, Sardesai, Schmiedekamp,
  Schmiedekamp, Schmitz, Schult, Shapiro-Albert, Siemens, Simon, Siwek, Stairs,
  Stinebring, Stovall, Sun, Susobhanan, Swiggum, Taylor, Taylor, Turner, Unal,
  Vallisneri, van Haasteren, Vigeland, Wahl, Wang, Witt, Young, \&
  Collaboration}]{Agazie2023}
Agazie, G., Anumarlapudi, A., Archibald, A.~M., {et~al.} 2023, The
  Astrophysical Journal Letters, 951, L8, \dodoi{10.3847/2041-8213/acdac6}

\bibitem[{{Andrews} {et~al.}(2018){Andrews}, {Huang}, {P{\'e}rez}, {Isella},
  {Dullemond}, {Kurtovic}, {Guzm{\'a}n}, {Carpenter}, {Wilner}, {Zhang}, {Zhu},
  {Birnstiel}, {Bai}, {Benisty}, {Hughes}, {{\"O}berg}, \&
  {Ricci}}]{andrews+2018}
{Andrews}, S.~M., {Huang}, J., {P{\'e}rez}, L.~M., {et~al.} 2018, \apjl, 869,
  L41, \dodoi{10.3847/2041-8213/aaf741}

\bibitem[{{Antoniadis} {et~al.}(2023{\natexlab{a}}){Antoniadis}, {Arumugam},
  {Arumugam}, {Babak}, {Bagchi}, {Bak Nielsen}, {Bassa}, {Bathula},
  {Berthereau}, {Bonetti}, {Bortolas}, {Brook}, {Burgay}, {Caballero},
  {Chalumeau}, {Champion}, {Chanlaridis}, {Chen}, {Cognard}, {Dandapat}, {Deb},
  {Desai}, {Desvignes}, {Dhanda-Batra}, {Dwivedi}, {Falxa}, {Ferdman},
  {Franchini}, {Gair}, {Goncharov}, {Gopakumar}, {Graikou}, {Grie{\ss}meier},
  {Guillemot}, {Guo}, {Gupta}, {Hisano}, {Hu}, {Iraci}, {Izquierdo-Villalba},
  {Jang}, {Jawor}, {Janssen}, {Jessner}, {Joshi}, {Kareem}, {Karuppusamy},
  {Keane}, {Keith}, {Kharbanda}, {Kikunaga}, {Kolhe}, {Kramer}, {Krishnakumar},
  {Lackeos}, {Lee}, {Liu}, {Liu}, {Lyne}, {McKee}, {Maan}, {Main},
  {Mickaliger}, {Nitu}, {Nobleson}, {Paladi}, {Parthasarathy}, {Perera},
  {Perrodin}, {Petiteau}, {Porayko}, {Possenti}, {Prabu}, {Quelquejay Leclere},
  {Rana}, {Samajdar}, {Sanidas}, {Sesana}, {Shaifullah}, {Singha}, {Speri},
  {Spiewak}, {Srivastava}, {Stappers}, {Surnis}, {Susarla}, {Susobhanan},
  {Takahashi}, {Tarafdar}, {Theureau}, {Tiburzi}, {van der Wateren}, {Vecchio},
  {Venkatraman Krishnan}, {Verbiest}, {Wang}, {Wang}, \&
  {Wu}}]{antoniadis+2023a}
{Antoniadis}, J., {Arumugam}, P., {Arumugam}, S., {et~al.} 2023{\natexlab{a}},
  arXiv e-prints, arXiv:2306.16214, \dodoi{10.48550/arXiv.2306.16214}

\bibitem[{{Antoniadis} {et~al.}(2023{\natexlab{b}}){Antoniadis}, {Arumugam},
  {Arumugam}, {Auclair}, {Babak}, {Bagchi}, {Bak Nielsen}, {Barausse}, {Bassa},
  {Bathula}, {Berthereau}, {Bonetti}, {Bortolas}, {Brook}, {Burgay},
  {Caballero}, {Caprini}, {Chalumeau}, {Champion}, {Chanlaridis}, {Chen},
  {Cognard}, {Crisostomi}, {Dandapat}, {Deb}, {Desai}, {Desvignes},
  {Dhanda-Batra}, {Dwivedi}, {Falxa}, {Fastidio}, {Ferdman}, {Franchini},
  {Gair}, {Goncharov}, {Gopakumar}, {Graikou}, {Grie{\ss}meier}, {Gualandris},
  {Guillemot}, {Guo}, {Gupta}, {Hisano}, {Hu}, {Iraci}, {Izquierdo-Villalba},
  {Jang}, {Jawor}, {Janssen}, {Jessner}, {Joshi}, {Kareem}, {Karuppusamy},
  {Keane}, {Keith}, {Kharbanda}, {Khizriev}, {Kikunaga}, {Kolhe}, {Kramer},
  {Krishnakumar}, {Lackeos}, {Lee}, {Liu}, {Liu}, {Lyne}, {McKee}, {Maan},
  {Main}, {Mickaliger}, {Middleton}, {Neronov}, {Nitu}, {Nobleson}, {Paladi},
  {Parthasarathy}, {Perera}, {Perrodin}, {Petiteau}, {Porayko}, {Possenti},
  {Prabu}, {Postnov}, {Quelquejay Leclere}, {Rana}, {Roper Pol}, {Samajdar},
  {Sanidas}, {Semikoz}, {Sesana}, {Shaifullah}, {Singha}, {Smarra}, {Speri},
  {Spiewak}, {Srivastava}, {Stappers}, {Steer}, {Surnis}, {Susarla},
  {Susobhanan}, {Takahashi}, {Tarafdar}, {Theureau}, {Tiburzi}, {Truant}, {van
  der Wateren}, {Valtolina}, {Vecchio}, {Venkatraman Krishnan}, {Verbiest},
  {Wang}, {Wang}, \& {Wu}}]{antoniadis+2023b}
---. 2023{\natexlab{b}}, arXiv e-prints, arXiv:2306.16227,
  \dodoi{10.48550/arXiv.2306.16227}

\bibitem[{{Arzoumanian} {et~al.}(2020){Arzoumanian}, {Baker}, {Brazier},
  {Brook}, {Burke-Spolaor}, {B{\'e}csy}, {Charisi}, {Chatterjee}, {Cordes},
  {Cornish}, {Crawford}, {Cromartie}, {Crowter}, {Decesar}, {Demorest},
  {Dolch}, {Elliott}, {Ellis}, {Ferdman}, {Ferrara}, {Fonseca},
  {Garver-Daniels}, {Gentile}, {Good}, {Hazboun}, {Islo}, {Jennings}, {Jones},
  {Kaiser}, {Kaplan}, {Kelley}, {Key}, {Lam}, {Lazio}, {Levin}, {Luo}, {Lynch},
  {Madison}, {McLaughlin}, {Mingarelli}, {Ng}, {Nice}, {Pennucci}, {Pol},
  {Ransom}, {Ray}, {Shapiro-Albert}, {Siemens}, {Simon}, {Spiewak}, {Stairs},
  {Stinebring}, {Stovall}, {Swiggum}, {Taylor}, {Vallisneri}, {Vigeland},
  {Witt}, {Zhu}, \& {NANOGrav Collaboration}}]{Arzoumanian2020}
{Arzoumanian}, Z., {Baker}, P.~T., {Brazier}, A., {et~al.} 2020, \apj, 900,
  102, \dodoi{10.3847/1538-4357/ababa1}

\bibitem[{{Avara} {et~al.}(2023){Avara}, {Krolik}, {Campanelli}, {Noble},
  {Bowen}, \& {Ryu}}]{avara+2023}
{Avara}, M.~J., {Krolik}, J.~H., {Campanelli}, M., {et~al.} 2023, arXiv
  e-prints, arXiv:2305.18538, \dodoi{10.48550/arXiv.2305.18538}

\bibitem[{{Barnes} \& {Hernquist}(1992)}]{Barnes1992}
{Barnes}, J.~E., \& {Hernquist}, L. 1992, \araa, 30, 705,
  \dodoi{10.1146/annurev.aa.30.090192.003421}

\bibitem[{{Begelman} {et~al.}(1980){Begelman}, {Blandford}, \&
  {Rees}}]{begelman+1980}
{Begelman}, M.~C., {Blandford}, R.~D., \& {Rees}, M.~J. 1980, \nat, 287, 307,
  \dodoi{10.1038/287307a0}

\bibitem[{{Boss}(2017)}]{boss2017}
{Boss}, A.~P. 2017, \apj, 836, 53, \dodoi{10.3847/1538-4357/836/1/53}

\bibitem[{{Bowen} {et~al.}(2017){Bowen}, {Campanelli}, {Krolik}, {Mewes}, \&
  {Noble}}]{bowen+2017}
{Bowen}, D.~B., {Campanelli}, M., {Krolik}, J.~H., {Mewes}, V., \& {Noble},
  S.~C. 2017, \apj, 838, 42, \dodoi{10.3847/1538-4357/aa63f3}

\bibitem[{{Bowen} {et~al.}(2019){Bowen}, {Mewes}, {Noble}, {Avara},
  {Campanelli}, \& {Krolik}}]{bowen_lump}
{Bowen}, D.~B., {Mewes}, V., {Noble}, S.~C., {et~al.} 2019, \apj, 879, 76,
  \dodoi{10.3847/1538-4357/ab2453}

\bibitem[{{Carr}(1980)}]{Carr1980}
{Carr}, B.~J. 1980, \aap, 89, 6.
\newblock \url{https://articles.adsabs.harvard.edu/pdf/1980A%26A....89....6C}

\bibitem[{{Charisi} {et~al.}(2016){Charisi}, {Bartos}, {Haiman},
  {Price-Whelan}, {Graham}, {Bellm}, {Laher}, \& {M{\'a}rka}}]{charisi+2016}
{Charisi}, M., {Bartos}, I., {Haiman}, Z., {et~al.} 2016, \mnras, 463, 2145,
  \dodoi{10.1093/mnras/stw1838}

\bibitem[{{Chen} {et~al.}(2020){Chen}, {Liu}, {Liao}, {Holgado}, {Guo},
  {Gruendl}, {Morganson}, {Shen}, {Zhang}, {Abbott}, {Aguena}, {Allam},
  {Avila}, {Bertin}, {Bhargava}, {Brooks}, {Burke}, {Carnero Rosell},
  {Carollo}, {Carrasco Kind}, {Carretero}, {Costanzi}, {da Costa}, {Davis}, {De
  Vicente}, {Desai}, {Diehl}, {Doel}, {Everett}, {Flaugher}, {Friedel},
  {Frieman}, {Garc{\'\i}a-Bellido}, {Gaztanaga}, {Glazebrook}, {Gruen},
  {Gutierrez}, {Hinton}, {Hollowood}, {James}, {Kim}, {Kuehn}, {Kuropatkin},
  {Lewis}, {Lidman}, {Lima}, {Maia}, {March}, {Marshall}, {Menanteau},
  {Miquel}, {Palmese}, {Paz-Chinch{\'o}n}, {Plazas}, {Sanchez}, {Schubnell},
  {Serrano}, {Sevilla-Noarbe}, {Smith}, {Suchyta}, {Swanson}, {Tarle},
  {Tucker}, {Norbert Varga}, \& {Walker}}]{chen+2020}
{Chen}, Y.-C., {Liu}, X., {Liao}, W.-T., {et~al.} 2020, \mnras, 499, 2245,
  \dodoi{10.1093/mnras/staa2957}

\bibitem[{{Chen} {et~al.}(2022){Chen}, {Zhai}, {Liu}, {Guo}, {Peng}, {Li},
  {SongSheng}, {Du}, {Hu}, \& {Wang}}]{chen+2022}
{Chen}, Y.-J., {Zhai}, S., {Liu}, J.-R., {et~al.} 2022, arXiv e-prints,
  arXiv:2206.11497, \dodoi{10.48550/arXiv.2206.11497}

\bibitem[{{Cimerman} {et~al.}(2023){Cimerman}, {Rafikov}, \&
  {Miranda}}]{cimerman+2023}
{Cimerman}, N.~P., {Rafikov}, R.~R., \& {Miranda}, R. 2023, arXiv e-prints,
  arXiv:2306.07341, \dodoi{10.48550/arXiv.2306.07341}

\bibitem[{{Combi} {et~al.}(2022){Combi}, {Lopez Armengol}, {Campanelli},
  {Noble}, {Avara}, {Krolik}, \& {Bowen}}]{combi+2021}
{Combi}, L., {Lopez Armengol}, F.~G., {Campanelli}, M., {et~al.} 2022, \apj,
  928, 187, \dodoi{10.3847/1538-4357/ac532a}

\bibitem[{{d'Ascoli} {et~al.}(2018){d'Ascoli}, {Noble}, {Bowen}, {Campanelli},
  {Krolik}, \& {Mewes}}]{dAscoli+2018}
{d'Ascoli}, S., {Noble}, S.~C., {Bowen}, D.~B., {et~al.} 2018, \apj, 865, 140,
  \dodoi{10.3847/1538-4357/aad8b4}

\bibitem[{{Dempsey} {et~al.}(2020){Dempsey}, {Mu{\~n}oz}, \&
  {Lithwick}}]{dempsey+2020}
{Dempsey}, A.~M., {Mu{\~n}oz}, D., \& {Lithwick}, Y. 2020, \apjl, 892, L29,
  \dodoi{10.3847/2041-8213/ab800e}

\bibitem[{{Dittmann} \& {Ryan}(2021)}]{Dittmann+2021}
{Dittmann}, A.~J., \& {Ryan}, G. 2021, \apj, 921, 71,
  \dodoi{10.3847/1538-4357/ac1bbd}

\bibitem[{{D'Orazio} {et~al.}(2016){D'Orazio}, {Haiman}, {Duffell},
  {MacFadyen}, \& {Farris}}]{Dorazio+2016}
{D'Orazio}, D.~J., {Haiman}, Z., {Duffell}, P., {MacFadyen}, A., \& {Farris},
  B. 2016, \mnras, 459, 2379, \dodoi{10.1093/mnras/stw792}

\bibitem[{{D'Orazio} {et~al.}(2013){D'Orazio}, {Haiman}, \&
  {MacFadyen}}]{dorazio+2013}
{D'Orazio}, D.~J., {Haiman}, Z., \& {MacFadyen}, A. 2013, \mnras, 436, 2997,
  \dodoi{10.1093/mnras/stt1787}

\bibitem[{{Duffell} {et~al.}(2020){Duffell}, {D'Orazio}, {Derdzinski},
  {Haiman}, {MacFadyen}, {Rosen}, \& {Zrake}}]{duffel:2020:massratio}
{Duffell}, P.~C., {D'Orazio}, D., {Derdzinski}, A., {et~al.} 2020, \apj, 901,
  25, \dodoi{10.3847/1538-4357/abab95}

\bibitem[{{Farris} {et~al.}(2014){Farris}, {Duffell}, {MacFadyen}, \&
  {Haiman}}]{Farris+2014}
{Farris}, B.~D., {Duffell}, P., {MacFadyen}, A.~I., \& {Haiman}, Z. 2014, \apj,
  783, 134, \dodoi{10.1088/0004-637X/783/2/134}

\bibitem[{{Farris} {et~al.}(2015){Farris}, {Duffell}, {MacFadyen}, \&
  {Haiman}}]{farris+2015b}
---. 2015, \mnras, 446, L36, \dodoi{10.1093/mnrasl/slu160}

\bibitem[{Favre(1969)}]{favre1969}
Favre, A. 1969, Problems of hydrodynamics and continuum mechanics, 231.
\newblock \url{https://cir.nii.ac.jp/crid/1571980075531969280}

\bibitem[{{Franchini} {et~al.}(2023){Franchini}, {Lupi}, {Sesana}, \&
  {Haiman}}]{franchini+2023}
{Franchini}, A., {Lupi}, A., {Sesana}, A., \& {Haiman}, Z. 2023, \mnras, 522,
  1569, \dodoi{10.1093/mnras/stad1070}

\bibitem[{{Franchini} \& {Martin}(2019)}]{franchini+2019}
{Franchini}, A., \& {Martin}, R.~G. 2019, \apjl, 881, L32,
  \dodoi{10.3847/2041-8213/ab3920}

\bibitem[{{Franchini} {et~al.}(2019){Franchini}, {Martin}, \&
  {Lubow}}]{franchini+2019b}
{Franchini}, A., {Martin}, R.~G., \& {Lubow}, S.~H. 2019, \mnras, 485, 315,
  \dodoi{10.1093/mnras/stz424}

\bibitem[{{Goodchild} \& {Ogilvie}(2006)}]{Goodchild+2006}
{Goodchild}, S., \& {Ogilvie}, G. 2006, \mnras, 368, 1123,
  \dodoi{10.1111/j.1365-2966.2006.10197.x}

\bibitem[{{Graham} {et~al.}(2015){Graham}, {Djorgovski}, {Stern}, {Drake},
  {Mahabal}, {Donalek}, {Glikman}, {Larson}, \& {Christensen}}]{graham+2015}
{Graham}, M.~J., {Djorgovski}, S.~G., {Stern}, D., {et~al.} 2015, \mnras, 453,
  1562, \dodoi{10.1093/mnras/stv1726}

\bibitem[{{Guti{\'e}rrez} {et~al.}(2021){Guti{\'e}rrez}, {Combi}, {Noble},
  {Campanelli}, {Krolik}, {L{\'o}pez Armengol}, \&
  {Garc{\'\i}a}}]{Gutierrez+2021}
{Guti{\'e}rrez}, E.~M., {Combi}, L., {Noble}, S.~C., {et~al.} 2021, arXiv
  e-prints, arXiv:2112.09773.
\newblock \doarXiv{2112.09773}

\bibitem[{{Haiman} {et~al.}(2009){Haiman}, {Kocsis}, \& {Menou}}]{Haiman:2009}
{Haiman}, Z., {Kocsis}, B., \& {Menou}, K. 2009, \apj, 700, 1952,
  \dodoi{10.1088/0004-637X/700/2/1952}

\bibitem[{Harris {et~al.}(2020)Harris, Millman, van~der Walt, Gommers,
  Virtanen, Cournapeau, Wieser, Taylor, Berg, Smith, Kern, Picus, Hoyer, van
  Kerkwijk, Brett, Haldane, del R{\'{i}}o, Wiebe, Peterson,
  G{\'{e}}rard-Marchant, Sheppard, Reddy, Weckesser, Abbasi, Gohlke, \&
  Oliphant}]{numpy}
Harris, C.~R., Millman, K.~J., van~der Walt, S.~J., {et~al.} 2020, Nature, 585,
  357, \dodoi{10.1038/s41586-020-2649-2}

\bibitem[{{Hawkins}(2002)}]{Hawkins2002}
{Hawkins}, M.~R.~S. 2002, \mnras, 329, 76,
  \dodoi{10.1046/j.1365-8711.2002.04939.x}

\bibitem[{{Huang} {et~al.}(2020){Huang}, {Andrews}, {Dullemond}, {{\"O}berg},
  {Qi}, {Zhu}, {Birnstiel}, {Carpenter}, {Isella}, {Mac{\'\i}as}, {McClure},
  {P{\'e}rez}, {Teague}, {Wilner}, \& {Zhang}}]{huang+2020}
{Huang}, J., {Andrews}, S.~M., {Dullemond}, C.~P., {et~al.} 2020, \apj, 891,
  48, \dodoi{10.3847/1538-4357/ab711e}

\bibitem[{Hunter(2007)}]{Hunter:2007}
Hunter, J.~D. 2007, Computing in Science \& Engineering, 9, 90,
  \dodoi{10.1109/MCSE.2007.55}

\bibitem[{{Hur{\'e}} \& {Pierens}(2009)}]{hure+2009}
{Hur{\'e}}, J.~M., \& {Pierens}, A. 2009, \aap, 507, 573,
  \dodoi{10.1051/0004-6361/200912348}

\bibitem[{Jones {et~al.}(2001--)Jones, Oliphant, Peterson, {et~al.}}]{scipy}
Jones, E., Oliphant, T., Peterson, P., {et~al.} 2001--, {SciPy}: Open source
  scientific tools for {Python}.
\newblock \url{http://www.scipy.org/}

\bibitem[{{Kato}(1983)}]{kato1983}
{Kato}, S. 1983, \pasj, 35, 249.
\newblock \url{https://articles.adsabs.harvard.edu/pdf/1983PASJ...35..249K}

\bibitem[{{Kley} {et~al.}(2008){Kley}, {Papaloizou}, \& {Ogilvie}}]{kley+2008}
{Kley}, W., {Papaloizou}, J.~C.~B., \& {Ogilvie}, G.~I. 2008, \aap, 487, 671,
  \dodoi{10.1051/0004-6361:200809953}

\bibitem[{{Komossa}(2006)}]{Komossa2006}
{Komossa}, S. 2006, \memsai, 77, 733.
\newblock \url{https://articles.adsabs.harvard.edu/pdf/2006MmSAI..77..733K}

\bibitem[{{Latter} \& {Ogilvie}(2006)}]{latter+2006}
{Latter}, H.~N., \& {Ogilvie}, G.~I. 2006, \mnras, 372, 1829,
  \dodoi{10.1111/j.1365-2966.2006.11014.x}

\bibitem[{{Li} {et~al.}(2023){Li}, {Wang}, \& {Zheng}}]{Li2023}
{Li}, J., {Wang}, Z., \& {Zheng}, D. 2023, \mnras, 522, 2928,
  \dodoi{10.1093/mnras/stad1168}

\bibitem[{{Lin} \& {Kratter}(2016)}]{lin+2016}
{Lin}, M.-K., \& {Kratter}, K.~M. 2016, \apj, 824, 91,
  \dodoi{10.3847/0004-637X/824/2/91}

\bibitem[{{Liu} {et~al.}(2019){Liu}, {Gezari}, {Ayers}, {Burgett}, {Chambers},
  {Hodapp}, {Huber}, {Kudritzki}, {Metcalfe}, {Tonry}, {Wainscoat}, \&
  {Waters}}]{liu+2019}
{Liu}, T., {Gezari}, S., {Ayers}, M., {et~al.} 2019, \apj, 884, 36,
  \dodoi{10.3847/1538-4357/ab40cb}

\bibitem[{{Liu} {et~al.}(2020){Liu}, {Koss}, {Blecha}, {Ricci}, {Trakhtenbrot},
  {Mushotzky}, {Harrison}, {Ichikawa}, {Kakkad}, {Oh}, {Powell}, {Privon},
  {Schawinski}, {Shimizu}, {Smith}, {Stern}, {Treister}, \& {Urry}}]{liu+2020}
{Liu}, T., {Koss}, M., {Blecha}, L., {et~al.} 2020, \apj, 896, 122,
  \dodoi{10.3847/1538-4357/ab952d}

\bibitem[{{Lubow}(1991)}]{lubow1991a}
{Lubow}, S.~H. 1991, \apj, 381, 259, \dodoi{10.1086/170647}

\bibitem[{{Lubow}(1994)}]{lubow1994}
---. 1994, \apj, 432, 224, \dodoi{10.1086/174563}

\bibitem[{{Lyra} {et~al.}(2016){Lyra}, {Richert}, {Boley}, {Turner}, {Mac Low},
  {Okuzumi}, \& {Flock}}]{lyra+2016}
{Lyra}, W., {Richert}, A. J.~W., {Boley}, A., {et~al.} 2016, \apj, 817, 102,
  \dodoi{10.3847/0004-637X/817/2/102}

\bibitem[{{Lyubarskij} {et~al.}(1994){Lyubarskij}, {Postnov}, \&
  {Prokhorov}}]{lyubarskij+1994}
{Lyubarskij}, Y.~E., {Postnov}, K.~A., \& {Prokhorov}, M.~E. 1994, \mnras, 266,
  583, \dodoi{10.1093/mnras/266.3.583}

\bibitem[{{MacFadyen} \& {Milosavljevi{\'c}}(2008)}]{macfadyen+2008}
{MacFadyen}, A.~I., \& {Milosavljevi{\'c}}, M. 2008, \apj, 672, 83,
  \dodoi{10.1086/523869}

\bibitem[{{Martin} {et~al.}(2014){Martin}, {Nixon}, {Lubow}, {Armitage},
  {Price}, {Do{\u{g}}an}, \& {King}}]{martin+2014}
{Martin}, R.~G., {Nixon}, C., {Lubow}, S.~H., {et~al.} 2014, \apjl, 792, L33,
  \dodoi{10.1088/2041-8205/792/2/L33}

\bibitem[{{Mu{\~n}oz} {et~al.}(2019){Mu{\~n}oz}, {Miranda}, \&
  {Lai}}]{Munoz2019}
{Mu{\~n}oz}, D.~J., {Miranda}, R., \& {Lai}, D. 2019, \apj, 871, 84,
  \dodoi{10.3847/1538-4357/aaf867}

\bibitem[{{Muley} {et~al.}(2021){Muley}, {Dong}, \& {Fung}}]{muley+2021}
{Muley}, D., {Dong}, R., \& {Fung}, J. 2021, \aj, 162, 129,
  \dodoi{10.3847/1538-3881/ac141f}

\bibitem[{{M{\"u}ller} {et~al.}(2012){M{\"u}ller}, {Kley}, \&
  {Meru}}]{Mueller+2012}
{M{\"u}ller}, T.~W.~A., {Kley}, W., \& {Meru}, F. 2012, \aap, 541, A123,
  \dodoi{10.1051/0004-6361/201118737}

\bibitem[{{Murray} {et~al.}(2000){Murray}, {Warner}, \&
  {Wickramasinghe}}]{murray+2000}
{Murray}, J.~R., {Warner}, B., \& {Wickramasinghe}, D.~T. 2000, \mnras, 315,
  707, \dodoi{10.1046/j.1365-8711.2000.03488.x}

\bibitem[{{Noble} {et~al.}(2012){Noble}, {Mundim}, {Nakano}, {Krolik},
  {Campanelli}, {Zlochower}, \& {Yunes}}]{noble+2012}
{Noble}, S.~C., {Mundim}, B.~C., {Nakano}, H., {et~al.} 2012, \apj, 755, 51,
  \dodoi{10.1088/0004-637X/755/1/51}

\bibitem[{{Ogilvie}(2001)}]{ogilvie2001}
{Ogilvie}, G.~I. 2001, \mnras, 325, 231,
  \dodoi{10.1046/j.1365-8711.2001.04416.x}

\bibitem[{Okuta {et~al.}(2017)Okuta, Unno, Nishino, Hido, \& Loomis}]{cupy}
Okuta, R., Unno, Y., Nishino, D., Hido, S., \& Loomis, C. 2017, in Proceedings
  of Workshop on Machine Learning Systems (LearningSys) in The Thirty-first
  Annual Conference on Neural Information Processing Systems (NIPS).
\newblock \url{http://learningsys.org/nips17/assets/papers/paper_16.pdf}

\bibitem[{{Oyang} {et~al.}(2021){Oyang}, {Jiang}, \& {Blaes}}]{oyang+2021}
{Oyang}, B., {Jiang}, Y.-F., \& {Blaes}, O. 2021, \mnras, 505, 1,
  \dodoi{10.1093/mnras/stab1212}

\bibitem[{{Panessa} {et~al.}(2019){Panessa}, {Baldi}, {Laor}, {Padovani},
  {Behar}, \& {McHardy}}]{Panessa2019}
{Panessa}, F., {Baldi}, R.~D., {Laor}, A., {et~al.} 2019, Nature Astronomy, 3,
  387, \dodoi{10.1038/s41550-019-0765-4}

\bibitem[{{Pe{\~n}il} {et~al.}(2022){Pe{\~n}il}, {Ajello}, {Buson},
  {Dom{\'\i}nguez}, {Westernacher-Schneider}, \& {Zrake}}]{penil+2022}
{Pe{\~n}il}, P., {Ajello}, M., {Buson}, S., {et~al.} 2022, arXiv e-prints,
  arXiv:2211.01894, \dodoi{10.48550/arXiv.2211.01894}

\bibitem[{{Pe{\~n}il} {et~al.}(2020){Pe{\~n}il}, {Dom{\'\i}nguez}, {Buson},
  {Ajello}, {Otero-Santos}, {Barrio}, {Nemmen}, {Cutini}, {Rani},
  {Franckowiak}, \& {Cavazzuti}}]{penil+2020}
{Pe{\~n}il}, P., {Dom{\'\i}nguez}, A., {Buson}, S., {et~al.} 2020, \apj, 896,
  134, \dodoi{10.3847/1538-4357/ab910d}

\bibitem[{{Plummer}(1911)}]{plummer1911}
{Plummer}, H.~C. 1911, \mnras, 71, 460, \dodoi{10.1093/mnras/71.5.460}

\bibitem[{{Rajagopal} \& {Romani}(1995)}]{Rajagopal1995}
{Rajagopal}, M., \& {Romani}, R.~W. 1995, \apj, 446, 543,
  \dodoi{10.1086/175813}

\bibitem[{{Reardon} {et~al.}(2023{\natexlab{a}}){Reardon}, {Zic}, {Shannon},
  {Hobbs}, {Bailes}, {Di Marco}, {Kapur}, {Rogers}, {Thrane}, {Askew}, {Bhat},
  {Cameron}, {Cury{\l}o}, {Coles}, {Dai}, {Goncharov}, {Kerr}, {Kulkarni},
  {Levin}, {Lower}, {Manchester}, {Mandow}, {Miles}, {Nathan}, {Os{\l}owski},
  {Russell}, {Spiewak}, {Zhang}, \& {Zhu}}]{reardon+2023a}
{Reardon}, D.~J., {Zic}, A., {Shannon}, R.~M., {et~al.} 2023{\natexlab{a}},
  \apjl, 951, L6, \dodoi{10.3847/2041-8213/acdd02}

\bibitem[{{Reardon} {et~al.}(2023{\natexlab{b}}){Reardon}, {Zic}, {Shannon},
  {Di Marco}, {Hobbs}, {Kapur}, {Lower}, {Mandow}, {Middleton}, {Miles},
  {Rogers}, {Askew}, {Bailes}, {Bhat}, {Cameron}, {Kerr}, {Kulkarni},
  {Manchester}, {Nathan}, {Russell}, {Os{\l}owski}, \& {Zhu}}]{reardon+2023b}
---. 2023{\natexlab{b}}, \apjl, 951, L7, \dodoi{10.3847/2041-8213/acdd03}

\bibitem[{{Rice} {et~al.}(2011){Rice}, {Armitage}, {Mamatsashvili}, {Lodato},
  \& {Clarke}}]{rice+2011}
{Rice}, W.~K.~M., {Armitage}, P.~J., {Mamatsashvili}, G.~R., {Lodato}, G., \&
  {Clarke}, C.~J. 2011, \mnras, 418, 1356,
  \dodoi{10.1111/j.1365-2966.2011.19586.x}

\bibitem[{{Roedig} {et~al.}(2012){Roedig}, {Sesana}, {Dotti}, {Cuadra},
  {Amaro-Seoane}, \& {Haardt}}]{Roedig+2012}
{Roedig}, C., {Sesana}, A., {Dotti}, M., {et~al.} 2012, \aap, 545, A127,
  \dodoi{10.1051/0004-6361/201219986}

\bibitem[{{Ryan} \& {MacFadyen}(2017)}]{Ryan+2017}
{Ryan}, G., \& {MacFadyen}, A. 2017, \apj, 835, 199,
  \dodoi{10.3847/1538-4357/835/2/199}

\bibitem[{{Sesana} {et~al.}(2012){Sesana}, {Roedig}, {Reynolds}, \&
  {Dotti}}]{sesana+2012}
{Sesana}, A., {Roedig}, C., {Reynolds}, M.~T., \& {Dotti}, M. 2012, \mnras,
  420, 860, \dodoi{10.1111/j.1365-2966.2011.20097.x}

\bibitem[{{Shi} \& {Krolik}(2015)}]{shi+2015}
{Shi}, J.-M., \& {Krolik}, J.~H. 2015, \apj, 807, 131,
  \dodoi{10.1088/0004-637X/807/2/131}

\bibitem[{{Sierra} {et~al.}(2021){Sierra}, {P{\'e}rez}, {Zhang}, {Law},
  {Guzm{\'a}n}, {Qi}, {Bosman}, {{\"O}berg}, {Andrews}, {Long}, {Teague},
  {Booth}, {Walsh}, {Wilner}, {M{\'e}nard}, {Cataldi}, {Czekala}, {Bae},
  {Huang}, {Bergner}, {Ilee}, {Benisty}, {Le Gal}, {Loomis}, {Tsukagoshi},
  {Liu}, {Yamato}, \& {Aikawa}}]{sierra+2021}
{Sierra}, A., {P{\'e}rez}, L.~M., {Zhang}, K., {et~al.} 2021, \apjs, 257, 14,
  \dodoi{10.3847/1538-4365/ac1431}

\bibitem[{{Sobacchi} {et~al.}(2017){Sobacchi}, {Sormani}, \&
  {Stamerra}}]{Sobacchi2017}
{Sobacchi}, E., {Sormani}, M.~C., \& {Stamerra}, A. 2017, \mnras, 465, 161,
  \dodoi{10.1093/mnras/stw2684}

\bibitem[{{Tart{\.{e}}nas} \& {Zubovas}(2020)}]{tartenas+2020}
{Tart{\.{e}}nas}, M., \& {Zubovas}, K. 2020, \mnras, 492, 603,
  \dodoi{10.1093/mnras/stz3484}

\bibitem[{{Tiede} {et~al.}(2020){Tiede}, {Zrake}, {MacFadyen}, \&
  {Haiman}}]{Tiede2020}
{Tiede}, C., {Zrake}, J., {MacFadyen}, A., \& {Haiman}, Z. 2020, \apj, 900, 43,
  \dodoi{10.3847/1538-4357/aba432}

\bibitem[{{Tiede} {et~al.}(2022){Tiede}, {Zrake}, {MacFadyen}, \&
  {Haiman}}]{Tiede2022}
---. 2022, \apj, 932, 24, \dodoi{10.3847/1538-4357/ac6c2b}

\bibitem[{{Wang} {et~al.}(2023){Wang}, {Bai}, {Lai}, \& {Lin}}]{wang+2023}
{Wang}, H.-Y., {Bai}, X.-N., {Lai}, D., \& {Lin}, D. N.~C. 2023, \mnras, 526,
  3570, \dodoi{10.1093/mnras/stad2884}

\bibitem[{{Westernacher-Schneider} {et~al.}(2022){Westernacher-Schneider},
  {Zrake}, {MacFadyen}, \& {Haiman}}]{WS2022}
{Westernacher-Schneider}, J.~R., {Zrake}, J., {MacFadyen}, A., \& {Haiman}, Z.
  2022, \prd, 106, 103010, \dodoi{10.1103/PhysRevD.106.103010}

\bibitem[{{Whitehurst} \& {King}(1991)}]{whitehurst+1991}
{Whitehurst}, R., \& {King}, A. 1991, \mnras, 249, 25,
  \dodoi{10.1093/mnras/249.1.25}

\bibitem[{{Xu} {et~al.}(2023){Xu}, {Chen}, {Guo}, {Jiang}, {Wang}, {Xu}, {Xue},
  {Nicolas Caballero}, {Yuan}, {Xu}, {Wang}, {Hao}, {Luo}, {Lee}, {Han},
  {Jiang}, {Shen}, {Wang}, {Wang}, {Xu}, {Wu}, {Manchester}, {Qian}, {Guan},
  {Huang}, {Sun}, \& {Zhu}}]{xu+2023}
{Xu}, H., {Chen}, S., {Guo}, Y., {et~al.} 2023, Research in Astronomy and
  Astrophysics, 23, 075024, \dodoi{10.1088/1674-4527/acdfa5}

\bibitem[{{Zrake} {et~al.}(2021){Zrake}, {Tiede}, {MacFadyen}, \&
  {Haiman}}]{Zrake:2021:eccentric}
{Zrake}, J., {Tiede}, C., {MacFadyen}, A., \& {Haiman}, Z. 2021, \apjl, 909,
  L13, \dodoi{10.3847/2041-8213/abdd1c}

\end{thebibliography}

\end{document}